\documentclass[11pt,a4paper]{article}
\usepackage{graphicx}
\usepackage{amssymb}
\usepackage{amsmath}
\usepackage{amsfonts}
\usepackage{dsfont}
\usepackage{mathtools}
\usepackage{array}
\usepackage{rotating}
\usepackage{bbold,amsfonts}
\allowdisplaybreaks

\usepackage[utf8]{inputenc}
\usepackage{bm}
\usepackage{xcolor}
\usepackage{float}
\usepackage{braket}
\usepackage{placeins}
\usepackage[height=8.8in,width=6.45in]{geometry}
\usepackage[font=small,labelfont=bf]{caption}
\usepackage[hidelinks]{hyperref}
\usepackage{booktabs,float,slashed}
\usepackage{standalone}
\usepackage{caption}
\usepackage{subcaption}
\usepackage{makecell}
\numberwithin{equation}{section}
\newcommand{\vev}[1]{{\left\langle #1 \right\rangle}}

\newcommand{\beq}{\begin{equation}}
\newcommand{\eeq}{\end{equation}}

\newcommand{\nord}[1]{:\! #1 \!:}

\DeclareMathOperator{\Tr}{Tr}
\DeclareMathOperator{\tr}{tr}

\newcommand{\ii}{\mathrm{i}}

\makeatletter
\newcommand*{\letterdef@}{}
\newcommand*{\letterdef}[3]{%
	\def\letterdef@##1{\expandafter\newcommand\csname #1\endcsname{#2{##1}}}%
	\@tfor\@tempa :=#3\do{\expandafter\letterdef@\expandafter{\@tempa}}}
\makeatother
\letterdef{c#1} {\mathcal}{ABCDEFGHIJKLMNOPQRSTUVWXYZ} % \cX = \mathcal{X}
\letterdef{rm#1}{\mathrm} {dDeimM} % \rmX = \mathrm{X} for X in {dDmM}

\DeclareMathAlphabet{\mathbb}{U}{msb}{m}{n} %math fonts like R^4 look much better

\DeclareMathAlphabet{\mathbb}{U}{msb}{m}{n} %math fonts like R^4 look much better

\begin{document}

\begin{titlepage}

\begin{flushright}
%\scriptsize 
{\small QMUL-PH-24-05}
\end{flushright}

\vspace*{10mm}
\begin{center}
{\LARGE \bf 
Exact results for giant graviton four-point correlators}

\vspace*{15mm}

{\Large Augustus Brown, Francesco Galvagno, and Congkao Wen}

\vspace*{8mm}
	
Centre for Theoretical Physics, Department of Physics and Astronomy, \\
Queen Mary University of London, London, E1 4NS, UK
			\vskip 0.3cm
			
	{\small
		E-mail:
		\texttt{a.a.x.brown,f.galvagno,c.wen@qmul.ac.uk}
	}
\vspace*{0.8cm}
\end{center}

\begin{abstract}

\vspace*{0.3cm}

We study the four-point correlator $\langle \mathcal{O}_2 \mathcal{O}_2 \mathcal{D} \mathcal{D} \rangle$ in $\mathcal{N}=4$ super Yang-Mills theory (SYM) with $SU(N)$ gauge group, where $\mathcal{O}_2$ represents the superconformal primary operator with dimension two, while $\mathcal{D}$ denotes a determinant operator of dimension $N$, which is holographically dual to a giant graviton D3-brane extending along $S^5$. We analyse the integrated correlator associated with this observable, obtained after integrating out the spacetime dependence over a supersymmetric invariant measure. Similarly to other classes of integrated correlators in $\mathcal{N}=4$ SYM, this integrated correlator can be computed through supersymmetric localisation on the four-sphere. Employing matrix-model recursive techniques, we demonstrate that the integrated correlator can be reformulated as an infinite sum of protected three-point functions with known coefficients. This insight allows us to circumvent the complexity associated with the dimension-$N$ determinant operator, significantly streamlining the large-$N$ expansion of the integrated correlator. In the planar limit and beyond, we derive exact results for the integrated correlator valid for all values of the 't Hooft coupling, and investigate the resurgent properties of their strong coupling expansion. Additionally, in the large-$N$ expansion with fixed (complexified) Yang-Mills coupling, we deduce the $SL(2, \mathbb{Z})$ completion of these results in terms of the non-holomorphic Eisenstein series. The proposed modular functions are confirmed by explicit instanton calculations from the matrix model, and agree with expectations from the holographic dual picture of known results in type IIB string theory.

\end{abstract}
\end{titlepage}
\setcounter{tocdepth}{2}
\tableofcontents

\section{Introduction}
\label{sec:intro}

Correlation functions of superconformal primary operators in 4d $\mathcal{N}=4$ super Yang-Mills theory (SYM) have played a major role in the context of Quantum Field Theories and holography. Their two- and three-point functions are known to be fully protected \cite{Lee:1998bxa}, while 
many relevant results have recently been obtained for the four-point function both in weak \cite{Eden:2012tu,Drummond:2013nda, Bourjaily:2016evz, Caron-Huot:2021usw} and strong coupling regime \cite{Goncalves:2014ffa,Rastelli:2016nze,Alday:2017xua,Caron-Huot:2018kta,Alday:2019nin,Drummond:2020dwr,Abl:2020dbx,Aprile:2020mus,Alday:2023mvu} (see \cite{Heslop:2022xgp} for a review with further references). A new approach for four-point correlators introduced by \cite{Binder:2019jwn} involves integrating the spacetime dependence over a supersymmetric invariant measure. The resulting integrated correlators can be computed exactly by supersymmetric localisation \cite{Pestun:2007rz} in terms of a matrix model on a four-sphere and provide powerful tools to go beyond the perturbative approach and to explore non-perturbative and exact properties of $\mathcal{N}=4$ SYM (see \cite{Dorigoni:2022iem} for a recent review on some of the developments about integrated correlators and the relevant references).

All these developments have been focusing on correlators of superconformal primaries with fixed conformal dimension $p$, which in general does not scale with $N$.\footnote{Integrated correlators of a special class of maximal multi-trace operators $(\cO_2)^p$ (and its similar generalisations) with general $p$ were studied in \cite{Paul:2023rka, Brown:2023why}. In particular, the explicit regimes $p\sim N$ and $p\sim N^2$ were considered for $(\cO_2)^p$ in \cite{Paul:2023rka}; and recently the geometry of the gravity dual of such a heavy maximal multi-trace operator has been constructed in \cite{Giusto:2024trt}. Despite their scaling, these operators are very different from the determinant operators that we are considering here. The integrated correlators with such operators obey some special recursion relations (\textit{i.e.} the Laplace-difference equations), and their large-$N$ properties are relatively more under control.} They transform in the $[0,p,0]$ representation of $SO(6)_R$ symmetry group of $\cN=4$ SYM, and can be written as a combination of traces of fundamental fields over the gauge algebra $\mathfrak{su}(N)$, see definitions \eqref{eq:mul-tr} and \eqref{eq:sin-tr}. In the holographic correspondence, these operators are dual to gravitons (for $p=2$) and higher Kaluza-Klein modes (for $p>2$) in type IIB string theory.   

In this paper we also study a different kind of half-BPS and gauge invariant operator, defined as determinants over the gauge algebra $\mathfrak{su}(N)$, see \eqref{eq:def_det} for a general definition and \eqref{eq:defD} for the definition of the maximal determinant operator. Determinant operators have been introduced in the AdS/CFT context \cite{Witten:1998xy,McGreevy:2000cw,Balasubramanian:2001nh,Berenstein:2002ke,Balasubramanian:2002sa,Berenstein:2003ah}, and have represented one of the most successful tests in holography. The maximal determinant operator has a protected conformal dimension, $\Delta_\cD = N$. Hence in the large-$N$ limit it is the ideal probe particle of AdS spacetime without deforming the geometry, and is known as the \textit{giant graviton}. For the case discussed in this paper, the dual picture of a pair of determinant operators is a D3-brane extended over an $S^3 \subset S^5$ direction and moving along AdS$_5$ geodesics in the dual type IIB string theory.  

From a field theory perspective, the interest in determinant operators lies in their baryonic-like behaviour, being gauge singlets composed of $N$ constituents. As heavy operators in planar $\cN=4$ SYM, they have represented the ideal playground to explore non-perturbative physics by using integrability techniques \cite{Berenstein:2005vf,Berenstein:2005fa,deMelloKoch:2007rqf,deMelloKoch:2007nbd,Bekker:2007ea}, which have mostly been used for exploring the spectrum of small deformations. However, due to the subtleties of heavy operators, far fewer results have been obtained for correlation functions involving determinant operators.  
Some classes of three-point functions with determinant operators have been studied in the literature, in particular in the half-BPS case the three-point correlator can be computed in the free field limit via Wick contractions \cite{Corley:2001zk,Bissi:2011dc,deMelloKoch:2004crq}, and several new techniques have been developed in \cite{Jiang:2019xdz,Jiang:2019zig} for the non-BPS sector, where integrability techniques have been combined with a worldsheet interpretation of three-point functions, in terms of overlaps of a boundary state with a closed string state. However, very little is known about the four-point functions involving determinant operators \cite{Jiang:2019xdz,Jiang:2019zig, Vescovi:2021fjf},
especially compared with the extensive set of results of four-point correlators of trace operators with fixed dimensions.

In this paper, we consider the correlator $\langle \mathcal{O}_2 \mathcal{O}_2 \mathcal{D} \mathcal{D}  \rangle$, where $\mathcal{O}_2$ is the superconformal primary operators of dimension two and $\mathcal{D}$ is the determinant operator. This observable has been studied in the weak 't Hooft coupling regime in \cite{Jiang:2019xdz,Jiang:2019zig} at one and two loops, and recently the same computation has been pushed to three-loop order \cite{Jiang:2023uut}. To explore new techniques to go beyond the perturbative approach, we show that the recent ideas of integrated correlators can also apply to this observable. Schematically we study the following observable (see section \ref{sec:2.2} for a proper definition):
\begin{equation}\label{eq:1.1}
    \mathcal{C}_{\cD}(\tau, \bar{\tau}; N)  =    \int   d \nu(x_i) \vev{\cO_2(x_1)\cO_2(x_2)\cD(x_3)\cD(x_4)}\,,
\end{equation}
where $\tau$ is the complexified Yang-Mills coupling and the precise form of the integration measure $d \nu(x_i)$ is given in \eqref{eq:measure}. This integrated correlator can be computed using a matrix model on $S^4$ via supersymmetric localisation. 
In particular, the double insertion of $\mathcal{O}_2$ is associated with the mass deformation of $\cN=4$ SYM, the so-called $\cN=2^*$ theory, whereas the insertions of determinant operators can be formally implemented as combinations of multitrace chiral primary operators with dimension $N$, after performing the Gram-Schmidt orthogonalisation procedure to unmix the operators between $\mathbb R^4$ and $S^4$ \cite{Baggio:2014sna,Baggio:2015vxa,Gerchkovitz:2016gxx}. 

As a first result, we compute the integrated correlator \eqref{eq:1.1} in the standard 't Hooft limit, with $\lambda = g^2_{{\rm YM}}N$ fixed and large $N$, where the instanton contributions are exponentially suppressed.
The standard large-$N$ matrix model techniques do not directly apply for insertions scaling with $N$, and the canonical Gram-Schmidt procedure for determinant operators when $N$ is very large appears to be very complicated. Hence we develop a special methodology, by rewriting the four-point integrated correlator as an infinite sum of BPS three-point functions in the presence of two determinant operators, which can be evaluated via free field theory results. The coefficients of the sum are evaluated through the recursive methods for Gaussian matrix models developed in \cite{Billo:2017glv,Billo:2018oog,Beccaria:2020hgy,Galvagno:2020cgq}.
Remarkably, we are able to obtain closed-form formulas for the weak coupling expansion of the integrated correlator to all orders in 't Hooft coupling in the planar limit and at next-to-leading order in $1/N$. More explicitly, the integrated correlator \eqref{eq:1.1} in the 't Hooft limit can be expressed in terms of the following topological expansion\,\footnote{The large-$N$ expansion for this observable is expressed in terms of powers of $1/N$, instead of the usual genus expansion in powers of $1/N^2$ for correlators of operators with fixed conformal dimensions. This is consistent with the fact that the determinant operators are dual to D3-branes which introduce boundaries of the worldsheet. Therefore the first two orders in the large-$N$ expansion $\mathcal{C}^{(0)}_{\cD}$ and $\mathcal{C}^{(1)}_{\cD}$ can be interpreted as the disc and annulus amplitudes in string theory, respectively.}
\begin{align}\label{eq:C_D_genusExp}
       \mathcal{C}_{\cD}(\lambda; N) = \sum_{g=0}^{\infty} N^{1-g} \, \mathcal{C}^{(g)}_{\cD}(\lambda) \, .
\end{align}
We find the leading large-$N$ expression for the integrated correlator is given by the following closed form expansion in $\lambda$:
\begin{align} \label{eq:C_D_genus0}
 \mathcal{C}^{(0)}_{\cD}(\lambda) = 4  \sum_{\ell=1}^{\infty}  (-1)^{\ell} {\zeta (2 \ell{+}1)   } \left[\binom{2 \ell+1}{\ell}^2-\binom{2 \ell+1}{\ell}\right] \left( {\lambda \over 16\pi^2}\right)^{\ell}  ~ , 
\end{align}
and the subleading order in the large-$N$ expansion takes a similar form:
\begin{align} \label{eq:C_D_genus1}
 \mathcal{C}^{(1)}_{\cD}(\lambda) = 2 \sum_{\ell=1}^\infty (-1)^{\ell-1} \zeta (2 \ell{+}1) \, (\ell{+}1) \left[ \binom{2 \ell+1}{\ell}^2 -(\ell+2) \binom{2\ell+1}{\ell}\right]\left(\frac{\lambda}{16 \pi^2} \right)^\ell ~ . 
\end{align}
These all-loop expressions represent the first important results of this paper. Then the perturbative results \eqref{eq:C_D_genus0} and \eqref{eq:C_D_genus1} can be resummed to \textit{exact functions} of the 't Hooft coupling $\lambda$, as shown in \eqref{eq:CD_exact} and \eqref{eq:CD_exact2}. Using Mellin-Barnes integrals, we further expand the exact formula in the strong coupling regime, and discuss the resurgence properties of the strong coupling expansion that can be interpreted as a sum over worldsheet instantons.

Next, we discuss the modular properties of our result, and its comparison with the expected structures from scattering amplitudes in the dual string theory. As expected for correlators of superconformal primaries in $\cN=4$ SYM, the four-point function $\langle \mathcal{O}_2 \mathcal{O}_2 \mathcal{D} \mathcal{D} \rangle$ is $SL(2, \mathbb{Z})$-invariant, due to the S-duality of $\cN=4$ SYM \cite{Montonen:1977sn}. This property can also be understood from the holographic interpretation at large $N$. Indeed, associating the boundary $\cD\cD$ pair insertion to a D3-brane in the bulk (see Figure \ref{fig:gravitonscattering} in section \ref{sec:sl2z}), we expect the four-point function to be $SL(2,\mathbb{Z})$ invariant, since D3-branes are self-dual under $SL(2,\mathbb{Z})$. Therefore, the integrated correlator \eqref{eq:1.1} must be described by modular functions. We justify this statement by studying the large-$N$ fixed-$\tau$ limit of the integrated correlator. After proposing the $SL(2, \mathbb{Z})$ completion of the results in terms of particular modular functions, the non-holomorphic Eisenstein series, we check this proposal against explicit instanton computations that are relevant in the large-$N$ fixed-$\tau$ limit, and we find a perfect agreement.
The proposed modular functions are also consistent with the expectations from flat space string amplitudes at finite string coupling \cite{Bachas:1999um, Green:2000ke, Basu:2008gt, Lin:2015ixa, Garousi:2011fc, Garousi:2011vs}.

The rest of the paper is structured as follows. In section \ref{sec:setup} we introduce the relevant half-BPS operators, distinguishing trace operators (with fixed dimensions) and determinant operators. In section \ref{sec:3}, we review the computation of integrated correlators from supersymmetric localisation, and we extend this definition to determinant operators. In section \ref{sec:4}, we lay out our method of computing the integrated correlator with determinant operators in the large-$N$ expansion. The final closed forms of the integrated correlator in the large-$N$ limit and beyond are displayed in section \ref{sec:5}, together with the strong coupling expansion and their resurgence properties. Finally in section \ref{sec:sl2z} we incorporate the non-perturbative instanton contributions and propose the $SL(2,\mathbb Z)$ completion of our result, which agrees with one-instanton computations as well as with the expectations from dual string theory scattering results. 
We conclude and comment on future research directions in section \ref{sec:7}.

The paper also includes five appendices. Appendix \ref{app:recursion} describes the formalism and the recursive techniques for the matrix model computation of integrated correlators. In appendix \ref{app:largeN-beta}, we apply the matrix model recursive techniques for a detailed computation of Gram-Schmidt orthogonalisation for trace operators, especially in the large-$N$ expansion. Appendix \ref{app:partial_contract} combines the partial contraction formula of determinant operators with matrix model techniques to compute the three-point coefficients.
Appendix \ref{app:2} describes the Gram-Schmidt orthogonalisation for the determinant operators and discusses their relation with the single-particle operators (SPO). Finally, appendix \ref{app:loops} discusses the comparison between our results for the integrated correlator and the field theory perturbative computation of the un-integrated correlator in the literature.

\section{Field theory set up and the integrated correlator}
\label{sec:setup}

\subsection{Trace and determinant operators: two- and three-point functions}
The first class of operators we introduce are the half-BPS trace operators with fixed conformal dimension in $\cN=4$ SYM with $SU(N)$ gauge group. These operators can be expressed in terms of multiple traces over the gauge algebra of the scalar fields $\Phi^I$ (with $I=1,\cdots, 6$), and take the following form:
\begin{align}  \label{eq:mul-tr}
\cO_{\vec p}\,(x, Y) = \frac{p_1 \dots p_m}{p}\cO_{[p_1]}(x, Y) \cdots \cO_{[p_m]}(x, Y) \, ,
\end{align} 
where each $\cO_{[p]}(x, Y)$ is a single-trace operator, 
\begin{align}  \label{eq:sin-tr}
\cO_{[p]}(x, Y) = \frac{1}{p}  Y_{I_1} \cdots Y_{I_p} \Tr\left( \Phi^{I_1}(x) \cdots \Phi^{I_p}(x) \right) \, ,
\end{align} 
and $Y_{I}$ is a null $SO(6)$ vector ($Y\cdot Y = 0$) that conveniently sums over the R-symmetry index $I$. Let us explain the notation in \eqref{eq:mul-tr}, that we use throughout the paper. We denote by $\vec p$ the vector $\vec p := [ p_1, \cdots, p_m ]$, $|\vec p|=m$ is the number of traces defining the operator, and $p= \sum_{i=1}^{m} p_i$ is the conformal dimension fixed by supersymmetry. 

These operators enjoy several properties, in particular their two- and three-point functions are protected by supersymmetry and do not receive any quantum corrections \cite{Lee:1998bxa}. In particular the two-point function of identical operators reads
\begin{align} \label{eq:OO_2point}
    \vev{ \cO_{\vec p}(x_1, Y_1)\cO_{\vec p}(x_2, Y_2)} =\cR_{\vec p}(N) (d_{12})^p \,,
\end{align}
where $d_{ij}$ is the free-field Wick contraction combining spacetime and R-symmetry vectors as
\begin{align}\label{eq:Rsymm_space_prop}
    d_{12}= \frac{Y_1\cdot Y_2}{x_{12}^2}\,,
\end{align}
and $\cR_{\vec p}(N)$ is an $N$-dependent normalization constant, which can be fixed arbitrarily. For a generic multitrace operator like \eqref{eq:mul-tr}, $\cR_{\vec p}(N)$ is conveniently fixed by free theory Wick contractions in the large-$N$ limit as
\begin{equation}\label{eq:2pt_normaliz}
    \cR_{\vec p}(N) = \frac{p!}{|\sigma_{\vec p}|} \left(N\right)^{p} +O(N^{p-2})\,,
\end{equation}
where $|\sigma_{\vec p}|$ is the size of the conjugacy class associated with the operator, see appendix \ref{app:2pt} for the explicit expression.

In this paper, we will also consider a different class of half-BPS operators, defined out of the determinant over gauge algebra $\mathfrak{su}(N)$ indices:
\begin{align}\label{eq:def_det}
    \det_k X = \frac{1}{k!} \epsilon_{i_1 \dots i_k \ell_1\dots \ell_{N-k}}\epsilon^{j_1 \dots j_k \ell_1\dots \ell_{N-k}} X_{j_1 \dots j_k}^{i_1 \dots i_k}\,,
\end{align}
where $i_n,j_n,\ell_n=1,\dots, N$ are fundamental indices of $\mathfrak{su}(N)$.
These operators can be considered the $\cN=4$ version of QCD baryons, being gauge singlets scaling with the number of colours. In particular, the main object of the present paper is the maximal determinant operator
\begin{align}\label{eq:defD}
    \cD(x, Y) = \det_N\,  \phi_I(x) Y^I \, , 
\end{align}
which has a protected conformal dimension $\Delta_{\cD}=N$. 
For finite and fixed values of $N$, $\cD(x, Y)$ can be expressed as a linear combination of the trace operators  $\cO_{\vec p}(x, Y) $ with dimension $p =N$. We will come back to this point in section \ref{sec:3.3} and further details can be found in appendix \ref{app:2}.

The property $\Delta_{\cD}=N$ becomes significantly problematic when considering the large-$N$ limit. From the four-dimensional field theory perspective the notion of planarity in the large-$N$ expansion is hard to define, and the large-$N$ properties of physical observables involving $\cD(x, Y)$ become much more subtle with respect to operators with fixed dimensions. From the holographic point of view, the determinant operators $\cD(x, Y)$ are dual to maximal giant gravitons, which are D3-branes wrapping an $S^3$ inside the $S^5$ of the AdS$_5 \times S^5$ supergravity background \cite{McGreevy:2000cw,Balasubramanian:2001nh}. 
Now we briefly review the results for correlation functions involving determinant operators that are needed for this paper.

The two-point function for determinant operators is protected by supersymmetry like \eqref{eq:OO_2point}, and is given by
\begin{align} \label{eq:DD_2point}
    \vev{ \cD(x_1, Y_1)\cD(x_2, Y_2)} =\cR_{\cD}(N) (d_{12})^N \,,
\end{align}
where $\cR_{\cD}(N)$ is the normalization constant, that can be computed by free field combinatorics \cite{Corley:2001zk} and scales as $\cR_{\cD}(N) \sim N!$. Higher-point functions involving only determinant operators have been computed in the free field limit \cite{deMelloKoch:2004crq,Kimura:2007wy} and at one loop \cite{Vescovi:2021fjf}. In this paper, we are interested in correlation functions involving both determinant operators and trace operators with fixed conformal dimensions.

The first example is the three-point function of two determinant operators and one half-BPS operator \eqref{eq:mul-tr} of dimension $p$, which is again protected by superconformal symmetry. The normalised three-point function is given as
\begin{align}\label{eq:3pt_DDO}
\frac{\vev{\cD(x_1,Y_1) \cD(x_2,Y_2) \cO_{\vec p}(x_3,Y_3)} }{\vev{\cD(x_1,Y_1) \cD(x_2,Y_2)}}=(\cR_{\vec p})^{\frac{1}{2}} \, \mathfrak{D}_{\vec p}\times \left(\frac{d_{23}d_{31}}{d_{12}}\right)^{\frac{p}{2}}\,,
\end{align}
where $\cR_{\vec p}$ is the two-point normalization given in \eqref{eq:2pt_normaliz}, $\mathfrak{D}_{\vec p}$ are the structure constants, and the last term corresponds to the spacetime and R-symmetry kinematic factor, which is fixed by superconformal symmetry. 
%\begin{equation}
%   (d_{12})^{N}\!\left(\frac{d_{23}d_{31}}{d_{12}}\right)^{\frac{p}{2}}=\frac{(Y_1\cdot Y_2)^{N-\frac{p}{2}}(Y_2\cdot Y_3)^{\frac{p}{2}}(Y_3\cdot Y_1)^{\frac{p}{2}}}{x_{12}^{2N-p}x_{23}^p x_{31}^p}\,.
%\end{equation}
The structure constants $\mathfrak{D}_{\vec p}$ only depend on combinatorics, and have been computed at large-$N$ for single-trace operators $\vec p = [p]$ in \cite{Jiang:2019xdz} using both field theory and a large-$N$ effective action approach (see for example their eq. (3.58)). For our purposes we define the normalised three-point coefficient $\mathfrak{C}_{\vec p}$ as
\begin{equation}\label{eq:C_p_def}
    \mathfrak{C}_{\vec p} (N) \equiv (\cR_{\vec p})^{\frac{1}{2}}\, \mathfrak{D}_{\vec p}\,.
\end{equation}
Starting from the result at leading order in $N$ for $\mathfrak{C}_{[p]}$ in \cite{Jiang:2019xdz}, we extend the computation of $\mathfrak{C}_{[p]}$ to the next order in the large-$N$ expansion, and we also compute the leading order at large $N$ for the multi-trace three-point coefficient $\mathfrak{C}_{\vec p}$. Both these results extending \cite{Jiang:2019xdz} will be relevant for our computations in section \ref{sec:5}. 

The normalised three-point coefficient for single-trace operators is given by
\begin{equation}\label{eq:3pt_FieldTh}
    \mathfrak{C}_{[p]} (N) = -\big(\ii^p+(-\ii)^p\big) N^{\frac{p}{2}} \left[1 - \frac{3p(p-2)}{8N} \right]+ O(N^{\frac{p}{2}-2})~ .
\end{equation}
The leading term was given in \cite{Jiang:2019xdz}, while the subleading contribution can be obtained in a similar way, by using the ``partially contracted giant graviton'' formula of \cite{Jiang:2019xdz} (see their eq. (4.7)). The idea is to express the partial Wick contraction of two determinant operators $\cD\cD$ in terms of trace operators, then to compute the two-point functions of trace operators.\footnote{The partial contraction formula in \cite{Jiang:2019xdz} is derived for $U(N)$ gauge group. However, as we show in appendix \ref{app:partial_contract}, the $SU(N)$ correction does not affect the leading and next-to-leading orders in large-$N$ that we consider in this paper.} 
We show how to compute the three-point coefficient and obtain the result \eqref{eq:3pt_FieldTh} with the help of Gaussian matrix model techniques in appendix \ref{app:partial_contract}.

From the same set of techniques we can compute $\mathfrak{C}_{\vec p}$ for the multi-trace operators at leading order in $N$:
\begin{align} \label{eq:multi-trace0}
   \mathfrak{C}_{\vec p} (N) =  \big(\ii^p+(-\ii)^p\big) (-1)^{m} \, N^{p\over 2}  + O(N^{\frac{p}{2}-1}) \, ,
\end{align}
where $p=\sum_{i=1}^m p_i$ with even $p_i$ and $p_i >0$. The result can be viewed as a product of the three-point functions of single-trace operators given in \eqref{eq:3pt_FieldTh}, as a result of the large-$N$ factorisation. For the goals of this paper, in the case of the multi-trace operators in \eqref{eq:multi-trace0}, we have only kept the leading-order contribution, and the cases with any odd $p_i$ are suppressed. 

These three-point coefficients \eqref{eq:3pt_FieldTh} and \eqref{eq:multi-trace0} are the crucial ingredients for computing the integrated four-point correlator, which we now introduce.

\subsection{Integrated correlators with determinant operators} \label{sec:2.2}
Let us now turn to the four-point function of interest, 
\begin{align}\label{eq:22DD_def}
    \vev{\cO_2(x_1, Y_1) \cO_2(x_2, Y_2) \cD(x_3, Y_3)\cD(x_4, Y_4)}  \,.
\end{align}
Because of the constraints from superconformal
symmetry, the correlator can be decomposed as follows \cite{Eden:2000bk, Nirschl:2004pa}: 
\begin{align} \label{eq:22pp}
     \vev{\cO_2(x_1, Y_1) \cO_2(x_2, Y_2) \cD(x_3, Y_3)\cD(x_4, Y_4)} = \mathcal{G}_{\rm free}(x_i, Y_i) + \mathcal{I}_4 (x_i, Y_i) \, d_{34}^{N-2} \, \mathcal{H}_{\cD}(U, V; \tau, \bar \tau) \, ,
\end{align}
where  $\mathcal{G}_{\rm free}(x_i, Y_i) $ is the free part, which may be computed by free-field Wick contractions, whereas the second term incorporates the quantum corrections. The full R-symmetry dependence can be factorized in the term $\cI_4 (x_i, Y_i)$ (and $d_{34}^{N-2}$), which is completely fixed by superconformal symmetry. Its explicit expression can be found in the literature, \textit{e.g.} in \cite{Paul:2022piq} (see their equation (2.11)), and is not needed for our purposes. The function $\mathcal{H}_{\cD} (U, V; \tau, \bar \tau)$ is the main focus of this paper and contains all the dynamics of this observable. It is a function of the Yang-Mills coupling
\begin{equation}\label{eq:tau_def}
    \tau:= \tau_1 + \ii\, \tau_2 = \frac{\theta}{2\pi} +\ii \frac{4\pi }{g_{\rm YM}^2}\,,
\end{equation}
and the cross ratios
\begin{align}
U= \frac{x_{12}^2 x_{34}^2 }{ x_{13}^2 x_{24}^2} \, , \quad \qquad V= \frac{x_{14}^2 x_{23}^2 }{ x_{13}^2 x_{24}^2} \, .
\end{align}
The function $\cH_\cD$ has been computed in perturbation theory at one and two loops in \cite{Jiang:2019xdz,Jiang:2019zig} and more recently at three loops in \cite{Jiang:2023uut}, see appendix \ref{app:loops} for a review of these achievements and a comparison with our result. No results beyond perturbation theory have been obtained for this observable yet.

In this paper, we apply the idea of integrated correlators, providing a fully non-perturbative method to study the determinant four-point function \eqref{eq:22DD_def}. This idea was originally derived for different classes of four-point functions, involving only trace operators. More specifically, we define the integrated correlator associated with 
$\mathcal{H}_{\cD}(U, V; \tau, \bar \tau)$ by integrating over the cross ratios with a certain measure \cite{Binder:2019jwn}\footnote{As shown in \cite{Wen:2022oky}, after factoring out some appropriate factors, the integral measure can also be written as \begin{align}\int \frac{d^4x_1 d^4x_2 d^4x_3 d^4 x_4}{{\rm vol}[SO(2,4)]} ~ ,\end{align} which is more convenient for evaluating the integrals. This is especially useful when the correlators are expressed in terms of conformal Feynman integrals in perturbation theory, then the integrated correlators can be viewed as a sum of periods of the Feynman integrals. We will utilise this approach when we compare our results with explicit weak coupling calculations of the four-point correlator in the literature in appendix \ref{app:loops}.} 
\begin{align} \label{eq:measure}
\cC_{\cD}(\tau, \bar \tau; N) = -\frac{2}{\pi} \int_0^{\infty} dr \int^{\pi}_0 d \theta {r^3 \sin^2 \theta \over U^2} \mathcal{H}_{\cD} (U, V; \tau, \bar \tau) \, , 
\end{align}
where $U=1-2r \cos\theta  +r^2$ and $V=r^2$, and we have made it clear that the result also depends on $N$ of gauge group $SU(N)$. The integration measure over the cross-ratios can be obtained by placing the $\cN=4$ theory on a $4$-sphere $S^4$ and deforming it by a mass parameter $m$ that preserves $\cN=2$ supersymmetry. The partition function of the resulting $\cN=2^*$ theory on the four-sphere can be computed via supersymmetric localisation in terms of a matrix model \cite{Pestun:2007rz}, so that the integrated correlator $\cC_{\cD}(\tau, \bar \tau; N)$ can be expressed in terms of a matrix model, as we now review in more detail.

\section{Integrated correlators as a matrix model}\label{sec:3}
\subsection{Review of integrated correlators for trace operators}
The integrated four-point function in $\cN=4$ SYM can be obtained from its massive deformation $\cN=2^*$ SYM.
By using the localisation techniques, the partition function of $\cN=2^*$ theory on a four-sphere can be described by a matrix model \cite{Pestun:2007rz}, in terms of an $N\times N$ Hermitian matrix $a$ taking values in the $\mathfrak{su}(N)$ gauge algebra.

Explicitly, the partition function can be written in terms of an integral over the eigenvalues of $a$ as follows:
\begin{equation}\label{eq:Z_matrixModel_tot}
    \begin{aligned}
\cZ(\tau, \tau'; m) =\! \int \! d\mu(a_i)  \left \vert \exp\bigg(\ii\, \pi \tau \sum_i a_i^2 +\ii \sum_{p>2} \pi^{p/2} \tau'_p \sum_{i} a_i^p \bigg)\right\vert^2 \!   
Z_{1\textrm{-loop}}(a; m) \left\vert Z_{\rm inst}(\tau, \tau', a;m) \right\vert^2\phantom{\bigg|} \, , 
\end{aligned}
\end{equation}
and we have defined
\begin{equation} \label{eq:measurea}
  d\mu(a_i)  =  \prod_{i=1}^N da_i \,\prod_{i<j} a_{ij}^2~  \delta\bigg( \sum_{i} a_i\bigg) \, ,
\end{equation}
where the integration variables $a_i$ are constrained by the $\mathfrak{su}(N)$ tracelessness condition $\sum_{i=1}^N a_i =0$. The exponential term contains the classical action proportional to the gauge coupling $\tau$, as well as the dependence on the higher dimensional couplings $\tau'_p$ which act as sources for chiral primary operators on the four-sphere; $Z_{1\textrm{-loop}}$ and $Z_{\rm inst}$ correspond to the perturbative one-loop determinant and the non-perturbative instanton contributions \cite{Nekrasov:2002qd}, respectively. In particular:
\begin{equation}\label{eq:Z1loop}
    Z_{1\textrm{-loop}}(a; m) = \frac{1}{H(m)^N} \prod_{i<j} \frac{H^2(a_{ij})}{H(a_{ij}+m)H(a_{ij}-m)}\,,
\end{equation}
and $H(x)$ is a product of Barnes G-functions $G(x)$:
\begin{equation}\label{eq:barnes}
    H(x) = \rme^{-(1+\gamma)x^2} G(1+\ii x) G(1-\ii x)\,.
\end{equation}
We postpone the introduction of the explicit expression for $Z_{\mathrm{inst}}$ to section \ref{sec:sl2z} when we consider $SL(2,\mathbb{Z})$ modular properties of the correlator.

For $m=\tau'_p=0$, both $Z_{1\textrm{-loop}}(a;m)$ and $ Z_{\rm inst}(\tau, \tau', a; m)$ reduce to $1$ and the matrix model becomes exactly Gaussian. In this case, the partition function reads
\begin{equation}\label{eq:Z_matrixModelGaussian}
\cZ_0\equiv \cZ(\tau, 0; 0) = \int d\mu(a_i)  \exp\bigg(\!\! -2\pi \tau_2\, \sum_i a_i^2\bigg)  \, , 
\end{equation}
and after rescaling the matrix $a$ as:
\begin{equation}\label{eq:tau_rescaling}
    a \rightarrow \frac{a}{\sqrt{2\pi \tau_2}}\,,
\end{equation}
one can write down any observable $f(a_i)$ evaluated in the Gaussian matrix model as:
\begin{equation}\label{eq:f_Gaussian}
    \vev{f(a_i)}_0 = \frac{1}{\cZ_0} \int   d\mu(a_i)  \exp\bigg(\!\! - \sum_i a_i^2\bigg) f(a_i) \, ,
\end{equation}
which in general will be some function of $N$.

We can now illustrate the procedure to compute integrated correlators from the $\cN=2^*$ partition function \eqref{eq:Z_matrixModel_tot}. As derived in \cite{Binder:2019jwn,Chester:2020dja}, taking multiple derivatives of \eqref{eq:Z_matrixModel_tot} is associated with the insertion of operators on the four-sphere. As an example, the quantity
\begin{align}\label{eq:2222_example1}
    \partial_m^4 \log \cZ(\tau, \tau'; m) \, {\vert}_{\tau',m=0} \, , 
\end{align}
corresponds to the integrated correlator of four $\cO_2$ operators, after using the idea that the mass deformation switches on a special R-symmetry channel (the so-called moment map operator) of the $\mathbf{20}^\prime$ operator in $\cN=4$ SYM. The localisation formula \eqref{eq:2222_example1} computes $\vev{\cO_2\cO_2\cO_2\cO_2}$ integrated over a different integration measure to \eqref{eq:measure} (it has an additional one-loop box integral). Further properties of this second class of integrated correlators can be found in \cite{Chester:2020dja, Chester:2020vyz, Alday:2023pet}. 

A similar idea holds when taking derivatives with respect to the mass $m$ as well as the couplings $\tau, \tau_p'$ and their complex conjugates.
As originally derived in \cite{Gerchkovitz:2016gxx}, taking derivatives of \eqref{eq:Z_matrixModel_tot} with respect to the couplings $\tau,\tau'_p$ and their complex conjugates is associated with the insertion of chiral/antichiral operators on the North/South poles of the four-sphere.
In particular $\partial_{\tau'_p}$ is associated to the insertion of the operator $O_p(a_i) = \tr a^p =  \sum_i a_i^p$. The action of multiple derivatives $\partial_{\tau'_{p_1,\dots, p_m}}:=\partial_{\tau'_{p_1}}\dots \partial_{\tau'_{p_m}}$ corresponds to the insertion of multitrace operators 
\begin{equation}\label{eq:def_multiTrace}
    O_{\vec p}(a) = O_{[p_1,\dots, p_m]} = \tr a^{p_1} \dots \tr a^{p_m}\,.
\end{equation}

However, the $S^4$ operators do not directly correspond to chiral primaries on $\mathbb R^4$. Due to the additional dimensionful parameter (\textit{i.e.} the radius of $S^4$), the operators with different dimensions on $S^4$ can mix, and in particular, unlike chiral operators on flat space, their two-point functions do not vanish. To create a correspondence between $S^4$ and $\mathbb R^4$, we need to perform a normal-ordering procedure, namely to impose orthogonality with respect to all the lower-dimensional (single and multi-trace) operators via the Gram-Schmidt procedure, as prescribed in \cite{Gerchkovitz:2016gxx,Rodriguez-Gomez:2016ijh,Billo:2017glv,Galvagno:2020cgq} in the context of extremal correlators in $\cN=2$ SCFTs (see also \cite{Binder:2019jwn, Paul:2022piq, Paul:2023rka, Brown:2023cpz, Brown:2023why} for the application of the Gram-Schmidt procedure to integrated correlators in $\cN=4$ SYM).
Therefore, given $O_{\vec p}(a)$ as defined in \eqref{eq:def_multiTrace} on the sphere, its normal ordered version (which is the proper matrix model version of the operators \eqref{eq:mul-tr} defined in $\mathbb R^4$) is given by\,\footnote{In this paper, we use straight capital letters for operators on $S^4$ and calligraphic letters for the corresponding normal ordered operators.}:
\begin{equation}
    \begin{split}
        \label{eq:On_NO}
   \cO_{\vec p}(a) \equiv \, \nord{O_{\vec p}(a)} =\cN_{\vec p} \left[ O_{\vec p}(a) +  \sum_{\vec q\,\vdash q< p}\alpha_{\vec p, \vec q}(N)~O_{\vec q}(a) \right] \, ,
    \end{split}
\end{equation}
where the sum in \eqref{eq:On_NO} runs over all the operators defined by the vector $\vec q = [q_1,\dots,q_n]$ with dimension $q=\sum_{i=1}^n q_i$, following the notation\,\footnote{Comparing with the previous literature \cite{Binder:2019jwn, Paul:2022piq, Paul:2023rka, Brown:2023cpz, Brown:2023why}, the coefficients $\alpha_{\vec p,\vec n}$ here correspond to $v_p^\mu$ there, with superscript $\mu$ running over all the operators with dimension lower than $p$. As we will see in the following, for our purposes it is convenient to stick to the notation $\alpha_{\vec p, \vec n}$ where the distinction between single and multi-trace operators is more evident.}
\begin{equation}\label{eq:sum_notation}
    \sum_{\vec q\,\vdash q} \equiv  \sum_{{\rm partitions~of~q}}\,.
\end{equation}
For $SU(N)$ we exclude all the partitions with any $q_i=1$, because this is associated with a dimension-one operator, which vanishes due to the tracelessness condition; therefore the partitions only include vectors with elements $q_i \geq 2$.  
We will use the shorthand notation \eqref{eq:sum_notation} throughout the paper.

Moreover, the coefficient $\cN_{\vec p}$ denotes a normalization factor, which can be fixed by computing the two-point functions, see appendix \ref{app:2pt}. Explicitly, we want the matrix model two-point function (computed in \eqref{eq:2pt_MM_multiT}) to be consistent with the field theory convention \eqref{eq:2pt_normaliz}:
\begin{align}
    \vev{\cO_{\vec p}(a) \cO_{\vec p}(a)}_0 = \cR_{\vec p}(N)\,,
\end{align}
which determines $\cN_{\vec p}$ to be 
\begin{equation}\label{eq:Normaliz_fixed}
   \cN_{\vec p} = 2^{\frac{p}{2}}\,.
\end{equation}

The coefficients $\alpha_{\vec p, \vec q}(N)$ are rational functions of $N$, which can be obtained by imposing for each $\vec p$ the following set of equations:
\begin{align}\label{eq:orth_procedure}
    \vev{\cO_{\vec p}(a) O_{\vec r}(a)}_0 =0\,,~~~ \forall r<p ~~~\text{and}~ \forall ~\text{partitions~of~}r \,,
\end{align}
where $\vev{\phantom x}_0$ stands for correlators in the Gaussian matrix model, as in \eqref{eq:f_Gaussian}. The orthogonalisation procedure \eqref{eq:orth_procedure} in general is rather involved; in appendix \ref{app:recursion} we describe a more refined way to compute the Gaussian integrals and to explicitly write down the normal ordered operators \eqref{eq:On_NO}.

Therefore, a general normalized integrated correlator (with the integration measure given in \eqref{eq:measure}) of two $\cO_2$ operators and two general chiral operators $\cO_{\vec p},\, \cO_{\vec q}$  with an equal dimension reads:\footnote{We will only focus on the cases where the two operators $\cO_{\vec p},\, \cO_{\vec q}$ are the same; in particular they will be the determinant operators. It is worth mentioning that when the operator $\cO_{\vec p}$ is not identical to $\cO_{\vec q}$, one needs to choose the normalisation factor $\sum\alpha_{\vec p,\vec r} \alpha_{\vec q,\vec s}\, \partial_{\tau'_{\vec m}} \partial_{\bar\tau'_{\vec s}}\log \cZ(\tau, \tau'; 0) \, {\vert}_{\tau'=0}$ in the denominator more carefully since it may vanish for some values of $\vec p$ and $\vec q$, and the choice of the normalisation also plays an important role for the integrated correlators to obey simple Laplace-difference equations \cite{Brown:2023cpz}.  } 
\begin{equation}\label{eq:22pp_example}
    \cC_{\vec p, \vec q}(\tau,\bar\tau;N) =    \dfrac{\displaystyle \sum_{\vec r \, ,\vec s}\alpha_{\vec p,\vec r} \, \alpha_{\vec q,\vec s} \,\partial_{\tau'_{\vec r}} \partial_{\bar\tau'_{\vec s}} \partial_m^2 \log \cZ(\tau, \tau'; m) \, {\vert}_{\tau',m=0}}{\displaystyle\sum_{\vec r \, ,\vec s}\alpha_{\vec p,\vec r} \, \alpha_{\vec q,\vec s} \,\partial_{\tau'_{\vec r}} \partial_{\bar\tau'_{\vec s}}\log \cZ(\tau, \tau'; 0) \, {\vert}_{\tau'=0}}\,. 
\end{equation}
 This localisation formula computes the integrated correlator as defined in \eqref{eq:measure}. The integrated correlators $\cC_{\vec p, \vec q}(\tau,\bar\tau; N)$ have been studied extensively in \cite{Brown:2023cpz, Paul:2023rka, Brown:2023why}, especially when the dimensions of the operators $\cO_{\vec p},\, \cO_{\vec q}$ are independent of $N$. 

\subsection{Normal ordered trace operators}
We now use the recursive techniques, originally developed in \cite{Billo:2017glv,Billo:2018oog,Beccaria:2020hgy,Galvagno:2020cgq} in the context of $\cN=2$ SCFTs and reviewed in Appendix \ref{app:recursion}, to evaluate Gaussian matrix model observables in an efficient way, and in particular to compute normal ordered operators as prescribed in \eqref{eq:On_NO} and \eqref{eq:orth_procedure}.

Starting from the Gaussian integral \eqref{eq:Z_matrixModelGaussian}, it is convenient to rewrite it in terms of a full $N\times N$ Hermitian matrix integral\,\footnote{The integral over the eigenvalues of $a$ is performed over a Cartan direction of $\mathfrak{su}(N)$ Lie algebra, hence the matrix integral over the full matrix $a$ is also referred as the full Lie algebra approach. As described by a direct comparison in \cite{Beccaria:2020hgy}, the full Lie algebra approach is more efficient for obtaining finite-$N$ results.}. Hence equations \eqref{eq:Z_matrixModelGaussian} and \eqref{eq:f_Gaussian} can be rewritten as follows:
\begin{align}\label{eq:Gauss_fullMatrix}
    \cZ_0=\int da~\rme^{-tr a^2}\,,~~~~~  \vev{f(a)}_0 = \int da~\rme^{-tr a^2} f(a)\,,
\end{align}
where the measure is normalised such that $\cZ_0=1$.

We use the matrix-model techniques to compute the mixing coefficients $\alpha_{\vec p, \vec n}$ as defined in \eqref{eq:On_NO}, using \eqref{eq:orth_procedure} for any given values of $p$. Using the Gaussian recursive formulas displayed in \eqref{eq:recursion_relat}, we write down the complete finite-$N$ coefficients up to very high values of $p$.  We explicitly write here an example for $\vec p = [4,2]$:
\begin{align}
    \cO_{[4,2]}\!=\!\cN_{[4,2]}\! \left[O_{[4,2]}{-} \frac{N^2{+}7}{2}O_{[4]}
{-}\frac{2N^2{-}3}{N} O_{[2,2]} {+} \frac{ 3(2N^2 {-} 3)(N^2 {+} 3)}{ 4N} O_{[2]} {-}\frac{(N^2 {-} 1)(N^2 {+} 3)(2 N^2 {-} 3)}{ 8N}\right],
\end{align}
where the last coefficient corresponds to the mixing with the identity operator. We refer to \eqref{NormalOrdExplicit} for other explicit formulas for normal ordered operators. 

For future convenience, we invert the relations \eqref{eq:On_NO}, and express all the multitrace operators on the sphere \eqref{eq:def_multiTrace} in terms of the normal ordered operators:
\begin{align}\label{multiT_to_NO}
    O_{\vec q}(a) =\widehat{\cO}_{\vec q}(a) +  \sum_{\vec p \,\vdash p<q}\beta_{\vec q, \vec p}(N)~\widehat{\cO}_{\vec p}(a)\,,
\end{align}
where we have defined the normalized normal ordered operator as:
\begin{equation}\label{eq:On_normalized}
    \widehat{\cO}_{\vec p}(a) \equiv \frac{\cO_{\vec p}(a)}{\cN_{\vec p}}\,.
\end{equation}
For example:
\begin{align}
 O_{[4,2]} \!=\widehat{\cO}_{[4,2]}\!+\! \frac{N^2\!+7}{2}\widehat{\cO}_{[4]}
\!+\!\frac{2N^2\!-3}{N} \widehat{\cO}_{[2,2]} \!+\! \frac{ 3(2N^2\!-3)(N^2\!+3)}{ 4N} \widehat{\cO}_{[2]}\!+\!\frac{(N^2\!-1)(N^2\!+3)(2 N^2\!-3)}{ 8N}\,.
\end{align}
Quite strikingly, the change of basis is highly symmetric, so that the coefficients $\alpha_{\vec q, \vec p}$ and $\beta_{\vec q, \vec p}$ are equal up to alternating signs:
\begin{align}\label{beta_to_alpha}
    \beta_{\vec q, \vec p}(N) = (-1)^{\frac{q-p}{2}} \alpha_{\vec q, \vec p}(N)\,.
\end{align}
This change of basis turns out to be crucial to evaluate the massive deformation in the matrix model, especially in the study of the large-$N$ limit of the integrated correlator in the presence of determinant operators which we will consider now.

%%%%%%%%%%%%%%%%%%%%%%%%%%%%%%%%%%%%%%%%%

\subsection{Determinant operators in the matrix model}\label{sec:3.3}
On the four-sphere, the determinant operator is defined  as a product over the eigenvalues of $a$:
\begin{equation}\label{eq:sphere_det}
    D(a_i) = \prod_{i=1}^N a_i \, . 
\end{equation}
As explained in the previous section, the integrated correlator should be defined for the corresponding normal ordered version $\cD(a_i)$, due to the mixing problem between $S^4$ and $\mathbb R^4$. In appendix \ref{app:2} we thoroughly explain the Gram-Schmidt procedure to uplift the determinant operator on the four-sphere \eqref{eq:sphere_det} to its normal ordered version $\cD(a_i)$ (we also comment on the relation between the determinant operator and the single-particle operator in the large-$N$ limit \cite{Aprile:2020uxk}). The explicit procedure described in appendix \ref{app:2} shows the difficulties in the Gram-Schmidt orthogonalisation of operators with dimension $\Delta=N$. 

In analogy with \eqref{eq:22pp_example}, we can formally define the matrix model version for the integrated correlator \eqref{eq:measure} in the presence of determinant operators $\cD$ as follows: 
\begin{align}\label{eq:C_D_def}
    \mathcal{C}_{\cD} (\tau, \bar{\tau};N) = \frac{\partial_{\mathcal{D}} \partial_{\mathcal{D}} \partial_m^2 \log \mathcal{Z}(\tau, \tau'; m) \, {\vert}_{\tau', m=0}}{\partial_{\mathcal{D}} \partial_{\mathcal{D}} \log \mathcal{Z}(\tau, \tau'; m) \, {\vert}_{\tau', m=0}} \, ,
\end{align}
where the notation $\partial_{\mathcal{D}}$ indicates the insertion of a determinant operator. More explicitly, for fixed values of $N$, as shown in \eqref{dettotraces}, we can express the determinant operator in terms of 
trace operators. Therefore, in principle, one may express the integrated correlator $\cC_{\cD}$ as a linear combination of $\cC_{{\vec p},{\vec q}}$ 
given in \eqref{eq:22pp_example}. However, computing $\cC_{\cD}$ in this way clearly becomes extremely difficult when $N$ is large, 
where the number of terms $\cC_{{\vec p},{\vec q}}$ in the linear combinations grows exponentially with $N$. 

To circumvent this complexity when computing the integrated correlator $\cC_\cD$ in 't Hooft limit,   we will follow a new methodology in order to bypass the Gram-Schmidt procedure on determinant operators. Following the definition \eqref{eq:C_D_def} we will insert $\cD(a_i)$ \textit{without explicitly performing the normal ordering}, and we will rather apply the Gram-Schmidt procedure on the mass deformed term. This reformulation represents the key point to carry out the computations of $\cC_\cD$ in the large-$N$ expansion, as described in the following sections.

\section{Integrated correlators as a sum of three-point functions}\label{sec:4}
In this section we focus on studying the integrated correlator $\cC_\cD$ \eqref{eq:C_D_def} from a perturbative approach of the matrix model. Here we are interested in the 't Hooft limit with large-$N$ and fixed-$\lambda$. In this regime, the instanton contribution is exponentially suppressed, so that we can simply drop the instanton partition function. The instanton effects will be discussed later in section \ref{sec:sl2z} when we consider the large-$N$ expansion of the integrated correlator with fixed-$\tau$ and study its modular properties.  

The goal of this section is to re-interpret the matrix model expression for $\cC_\cD$ in order to avoid explicitly performing the Gram-Schmidt orthogonalization for the determinant operators and to express the matrix-model expression of the integrated correlator in terms of three-point functions on $\mathbb{R}^4$. Let us outline the strategy. After expanding $Z_{\textrm{1-loop}}$ perturbatively, we use the inverse Gram-Schmidt procedure \eqref{multiT_to_NO} to rewrite $Z_{\textrm{1-loop}}$ purely in terms of normal ordered operators. 
The final outcome for the integrated correlator $\cC_{\cD}$ is an infinite sum of BPS and normal-ordered three-point functions in the presence of determinant operators, which can be evaluated using the explicit $\mathbb R^4$ planar results \eqref{eq:3pt_FieldTh}.\footnote{The procedure can be applied to any other integrated correlators of the form given in \eqref{eq:22pp_example};  here, we will only focus on the integrated correlator involving the determinant operators, for which some other approaches become difficult, as we have emphasised.} This method is powerful enough to derive closed-form formulas for the perturbative expansion for the leading and subleading orders of large-$N$ expansion.

\subsection{Mass deformation as an insertion in Gaussian matrix model}
Let us rewrite the partition function \eqref{eq:Z_matrixModel_tot} for $\tau'_p=0$ in the zero-instanton sector ($|Z_{\mathrm{inst}} |^2=1$) as the following integral over the matrix $a$, rescaled as in \eqref{eq:tau_rescaling}
\begin{align}
   \cZ(m,g_{\rm YM}^{2})=\int \!da~\rme^{-\tr\,a^2}\,Z_{\textrm{1-loop}}(a;m,g_{\rm YM}^{2})\,.
\label{Z2*}
\end{align} 
By expanding the Barnes G-functions \eqref{eq:barnes}, it is possible to rephrase the 1-loop term \eqref{eq:Z1loop} in terms of traces on the matrix $a$, and in the limit of small $m$ we can expand it perturbatively in $g_{\rm YM}^2$ \cite{Russo:2013kea}: 
\begin{align}
\log Z_{\textrm{1-loop}}=-m^2 \bigg[\sum_{\ell=1}^\infty\sum_{j=0}^{2\ell}(-1)^{\ell+j}\left(\frac{g_{\rm YM}^2}{8\pi^2}\right)^\ell\binom{2\ell}{j}\,(2\ell+1)\zeta(2\ell+1)\tr\,a^{2\ell-j}\,\tr\,a^{j}\bigg]+O(m^4)\,.
\end{align} 
At leading order in the mass, we can view the mass deformation as an insertion in the Gaussian matrix model, so that we rewrite the partition function as
\begin{align}\label{d2mZ_inMM}
  \partial_m^2\log \cZ(m,g_{\rm YM}^2)\Big|_{m=0}=  
  \int \!da~\rme^{-\tr\,a^2}\, \mathbf{M}_2(a,g_{\rm YM}^2)
\,\equiv\,\vev{\mathbf{M}_2(a,g_{\rm YM}^2)}_0\,,
\end{align}
where $\mathbf{M}_2$ can be seen as an infinite sum of operators on the sphere
\begin{align}\label{M2_doubleTraces}
\mathbf{M}_2(a,g_{\rm YM}^2)=-2\sum_{\ell=1}^\infty\sum_{j=0}^{2\ell}(-1)^{j+\ell} (2\ell+1) \binom{2\ell}{j}\,\zeta(2\ell+1)\Big(\frac{g_{\rm YM}^2}{8\pi^2}\Big)^\ell O_{[2\ell-j,j]}\,.
\end{align}
It is now useful to change basis and rewrite $\mathbf{M}_2$
in terms of the normal-ordered operators $\cO_{\vec p}(a)$, by following the transformation rule in \eqref{multiT_to_NO} and the definition \eqref{eq:On_normalized}:
\begin{align}\label{M2_multiTraces}
\mathbf{M}_2(a,g_{\rm YM}^2)=\!-2\sum_{\ell=1}^\infty\sum_{j=0}^{2\ell}(-1)^{j+\ell} (2\ell+1) \binom{2\ell}{j}\,\zeta(2\ell+1)\Big(\frac{g_{\rm YM}^2}{8\pi^2}\Big)^\ell \! \left[\widehat{\cO}_{[2\ell-j,j]} +\!\! \sum_{\vec p\,\vdash p< 2\ell}\beta_{[2\ell-j,j], \vec p}~\widehat{\cO}_{\vec p}\right]\,.
\end{align}
Let us stress the idea behind this rewriting: we trade the insertion of the operator \eqref{M2_doubleTraces} containing only double traces (but with no direct physical meaning on $\mathbb R^4$) with the operator \eqref{M2_multiTraces} written in terms of all the normal ordered multi-trace operators, which are the matrix model equivalent of the $\mathbb R^4$ operators.

\subsection{Integrated correlator with determinant operators}\label{sec:4_2}
After inserting the massive deformation in the Gaussian matrix model as in \eqref{d2mZ_inMM} we can rewrite the integrated correlator in the presence of determinant operators \eqref{eq:C_D_def} as follows:
\begin{align}\label{eq:CD_as_vevM2}
    \mathcal{C}_{\cD}  (g^2_{\rm YM}; N) = \frac{\vev{\mathcal{D}(a) \mathcal{D}(a) \mathbf{M}_2(a,g_{\rm YM}^2)}_0 }{\vev{\mathcal{D}(a) \mathcal{D}(a)}_0}- \vev{\mathbf{M}_2(a,g_{\rm YM}^2)}_0 \equiv \vev{\mathbf{M}_2(a,g_{\rm YM}^2)}_{\cD}\,,
\end{align}
where we defined the notation $\vev{\phantom{|}}_{\cD}$ to identify the connected normalized three point function in the presence of determinant operators. Using  \eqref{M2_multiTraces}, we get: 
\begin{equation}\label{CN_3pt}
    \begin{split}
    \mathcal{C}_{\cD}(g^2_{\rm YM}; N) = \, -2\sum_{\ell=1}^\infty\sum_{j=0}^{2\ell}&(-1)^{j+\ell} (2\ell+1) \binom{2\ell}{j}\,\zeta(2\ell+1)\Big(\frac{g_{\rm YM}^2}{8\pi^2}\Big)^\ell \\ &
\left(\!\vev{\widehat{\cO}_{[2\ell-j,j]}}_\cD +\! \sum_{\vec p \,\vdash p<2\ell}\beta_{[2\ell-j,j], \vec p}~\vev{\widehat{\cO}_{\vec p}}_\cD\!\right) \, .
\end{split}
\end{equation}
Note that the sum over $\vec p$ does not include the identity (so $p>0$), because this would provide a disconnected contribution which vanishes in the definition \eqref{eq:CD_as_vevM2}. 

We have now reduced the matrix model computation for the integrated correlator in the presence of determinant operators to an infinite sum over three-point functions of normal ordered operators. We will now study the 't Hooft large-$N$ expansion of the integrated correlator utilising the large-$N$ expansion of the three-point functions and the properties of $\beta$-coefficients. 

\subsubsection{Large-$N$ expansion}

Since all the operators are now normal ordered, we can directly apply the large-$N$ field theory results of the three-point functions and the $\beta$-coefficients that appear in  \eqref{CN_3pt}. In order to calculate the integrated correlator at leading and subleading order in $N$, we expand both the three-point functions and the $\beta$-coefficients in the large-$N$ limit. We find that, in general, the three-point functions behave as   
\begin{align} \label{eq:expansion1}
    \vev{\widehat{\cO}_{\vec{p}}}_\cD = \vev{\widehat{\cO}_{\vec{p}}}_\cD^{(0)} N^{\frac{p}{2}} + \vev{\widehat{\cO}_{\vec{p}}}_\cD^{(1)} N^{\frac{p}{2}-1} + \ldots\, ,
    \end{align}
where all the $p_i$ in $\vec p$ are even and $p_i>0$. From the field theory result \eqref{eq:3pt_FieldTh} we can read the coefficients for the single-trace operators which are relevant to our computation:
\begin{align} \label{eq:3ptOhat}
\vev{\widehat{\cO}_{[2p]}}_\cD^{(0)} =(-1)^{p-1}\, 2^{1-p}  \, , \qquad \vev{\widehat{\cO}_{[2p]}}_\cD^{(1)} =3 p(p-1)(-1)^{p} 2^{-p}\, ,
\end{align} 
where we have also taken into account the normalisation factor \eqref{eq:Normaliz_fixed}. For multi-trace operators we only need the leading-order contribution shown in \eqref{eq:multi-trace0}, and after including the normalisation factor \eqref{eq:Normaliz_fixed} we find: 
\begin{align} \label{eq:multi-trace}
    \vev{ \widehat{\cO}_{[2p_1, \ldots , 2p_m]}}_\cD^{(0)} = 2^{1-p} \, (-1)^{p-m} \, ,
\end{align}
where again $p=\sum_{i=1}^m p_i$ and $p_i >0$. In all the expressions \eqref{eq:expansion1}, \eqref{eq:3ptOhat} and \eqref{eq:multi-trace}, we have required all the elements in $\vec{p}$ to not be $0$. As shown in \eqref{tnodd}, each `$0$' in $\vec{p}$ would contribute with an additional factor of $N$ to \eqref{eq:expansion1}. For instance, in \eqref{CN_3pt} the $j=0$ term generates a contribution $ \vev{\widehat{\cO}_{[2\ell,0]}}_\cD$ that can be evaluated as follows:
\begin{align} \label{eq:expansion3}
     \vev{\widehat{\cO}_{[2\ell,0]}}_\cD =N  \vev{\widehat{\cO}_{[2\ell]}}_\cD= \vev{\widehat{\cO}_{[2\ell]}}_\cD^{(0)} N^{\ell+1} + \vev{\widehat{\cO}_{[2\ell]}}_\cD^{(1)} N^{\ell} + \ldots\, .
\end{align}  

The other ingredients that determine the scaling with $N$ are the mixing coefficients.
The large-$N$ properties of $\beta_{\vec{q},\vec{p}}$ are analysed in appendix \ref{app:largeN-beta}. From many specific examples obtained from the general recursive techniques of the matrix model, we find they behave as:  
    \begin{align}
    \beta_{\vec{q},\vec{p}} = \beta_{\vec{q},\vec{p}}^{(0)} \, N^{\frac{q-p}{2}+ n-m} + \beta_{\vec{q},\vec{p}}^{(1)}\, N^{\frac{q-p}{2}+ n-m-2} + \ldots \, , \label{eq:expansion2}
\end{align}
where  $p=\sum_{i=1}^m p_i$ and $q=\sum_{i=1}^n q_i$, and $m$ and $n$ are the lengths of the vectors $\vec{p}$ and $\vec{q}$ respectively. Furthermore, all the elements in  $\vec{q}$ and $\vec{p}$ are even, since $\langle\widehat{\cO}_{\vec{p}}\rangle_\cD$ and $\beta_{\vec{q},\vec{p}}$ with any odd terms are suppressed in large-$N$ to the order we consider. See appendix \ref{app:largeN-beta} for the explicit expressions of $\beta_{\vec{q},\vec{p}}$ that are used for the present analysis.

Let us remark the difference between the large-$N$ expansions of $\langle\widehat{\cO}_{\vec{p}}\rangle_\cD$ in \eqref{eq:expansion1} and the coefficients $\beta_{\vec{q},\vec{p}}$ in \eqref{eq:expansion2}. The expansion for $\langle\widehat{\cO}_{\vec{p}}\rangle_\cD$ is organised in powers of $1/N$, whereas the expansion parameter of $\beta_{\vec{q},\vec{p}}$ is $1/N^2$. This is expected and has also a deeper meaning. The determinant operators are dual to D3-branes introducing boundaries for open strings, and the three-point coefficient \eqref{eq:expansion1} can be seen as overlaps with boundary states. On the other hand, the mixing coefficients $\beta_{\vec{q},\vec{p}}$ come from two-point functions of half-BPS operators with fixed dimensions, which are dual to the closed-string KK modes in type IIB string theory. So their expansion in the large-$N$ limit follows the usual genus expansion in powers of $1/N^2$. 

Using \eqref{eq:expansion1}, \eqref{eq:expansion2} and \eqref{eq:expansion3}, we can therefore expand the terms in \eqref{CN_3pt} in the large-$N$ limit. We see that the weak-coupling expansion gets naturally organised in terms of the 't Hooft coupling $\lambda=N g^2_{\rm YM}$. Collecting the leading and subleading terms in the large-$N$ limit, we find 
\begin{align} \label{cDlead+sub}
    \mathcal{C}_{\cD}(\lambda; N) &=  -2\sum_{\ell=1}^\infty (-1)^{\ell} (2\ell+1)\,\zeta(2\ell+1)\Big(\frac{\lambda}{8\pi^2}\Big)^\ell\, \times \\
    &\Bigg( N  \Bigg[ 2\vev{ \widehat{\cO}_{[2\ell]}}_\cD^{(0)} \, + \, \sum_{j=0}^{\ell} \binom{2\ell}{2j} \sum_{p=1}^{\ell-1} \beta_{[2\ell-2j,2j],[2p]}^{(0)} \vev{\widehat{\cO}_{[2p]}}_\cD^{(0)} \Bigg]  \cr
    &+ \Bigg[ 2\vev{ \widehat{\cO}_{[2\ell]}}_\cD^{(1)} \, + \sum_{j=0}^{\ell}\binom{2\ell}{2j} \sum_{p=1}^{\ell-1} \beta_{[2\ell-2j,2j],[2p]}^{(0)} \vev{ \widehat{\cO}_{[2p]}}_\cD^{(1)}   + \sum_{j=1}^{\ell-1} \binom{2\ell}{2j} \vev{ \widehat{\cO}_{[2\ell-2j,2j]}}_\cD^{(0)} \cr 
    &+ \sum_{j=0}^{\ell}\binom{2\ell}{2j} \sum_{p=2}^{\ell-1} \sum_{p_1=1}^{\lfloor {p\over 2} \rfloor} \beta_{[2\ell-2j,2j],[2 p_1, 2p-2p_1]}^{(0)} \vev{ \widehat{\cO}_{[2p_1, 2p-2p_1]}}_\cD^{(0)} \Bigg]  \Bigg) + O(N^{-1}) \, , \nonumber
\end{align}
where we have kept the leading and subleading terms and omitted the higher order terms. In particular, the coefficients $\beta_{\vec{q},\vec{p}}^{(1)}$ in \eqref{eq:expansion2} do not contribute since they differ from the leading-order term by $1/N^2$.

In the next section, we will examine each order in $N$ separately using the explicit expressions of the three-point functions $\vev{\widehat{\cO}_{\vec{k}}}_\cD$ and the coefficients $\beta_{\vec{q},\vec{p}}^{(0)}\,$. This analysis will allow us to obtain exact expressions as functions of 't Hooft coupling $\lambda$ for the integrated correlator $\mathcal{C}_{\cD}(\lambda; N)$ in the large-$N$ expansion. 

\section{Exact results at weak and strong coupling} \label{sec:5}

In this section, we consider the large-$N$ 't Hooft expansion of the integrated correlator. As we can see from \eqref{cDlead+sub}, $\cC_\cD$ is organised in terms of the following topological expansion:
\begin{align}\label{eq:C_D_genus}
       \mathcal{C}_{\cD}(\lambda; N) = \sum_{g=0}^{\infty} N^{1-g} \, \mathcal{C}^{(g)}_{\cD}(\lambda) \, . 
\end{align}
As we mentioned in the introduction, the large-$N$ expansion for this observable is decomposed in powers of $1/N$, instead of the usual $1/N^2$ for correlators of trace operators with fixed conformal dimensions. This is consistent with the holographic interpretation that the determinant operators are dual to D3-branes in the string theory. In this paper we will concentrate on the Leading order $\mathcal{C}^{(0)}_{\cD}$ and Next-to-Leading order $\mathcal{C}^{(1)}_{\cD}$ in the expansion \eqref{eq:C_D_genus}, which in the dual string theory interpretation correspond to the disc string amplitude and the annulus amplitude respectively. We will come back to this interpretation in section \ref{sec:sl2z}.

\subsection{Weak coupling results in large-$N$ expansion}

Using the ingredients discussed in the previous section, we now derive the all-order expression for the large-$N$ expansion of the integrated correlator $\mathcal{C}_{\cD}(\lambda; N)$ in the leading planar limit and the next-to-leading correction. We then resum the perturbative results to obtain exact expressions for $\mathcal{C}_{\cD}(\lambda; N)$, valid for any `t Hooft coupling $\lambda$, and we analyse the strong coupling expansion and the associated resurgent properties. 

\subsubsection{Leading-$N$ expression and comparison with field theory results}

We begin with the leading-$N$ contribution. From \eqref{cDlead+sub}, the leading-$N$ term can be written as 
\begin{align}\label{cleadingN}
    \begin{split}
        \mathcal{C}^{(0)}_{\cD}(\lambda) &= -2 \sum_{\ell=1}^\infty (-1)^{\ell} (2\ell+1)\,\zeta(2\ell+1)\Big(\frac{\lambda}{8\pi^2}\Big)^\ell \cr
    &\Bigg[ 2\vev{ \widehat{\cO}_{[2\ell]}}_\cD^{(0)} \, + \, \sum_{p=1}^{\ell-1} \vev{\widehat{\cO}_{[2p]}}_\cD^{(0)} \sum_{j=0}^{\ell} \binom{2\ell}{2j} \beta_{[2\ell-2j,2j],[2p]}^{(0)}  \Bigg]\,.
    \end{split}
\end{align}
As discussed in appendix \ref{app:largeN-beta}, we find that the coefficients $\beta_{[2l-2j,2j],[2p]}^{(0)}$ are given by 
\begin{align} \label{eq:beta0}
\beta_{[2\ell-2j,2j], [2p] }^{(0)} = 2^{p-\ell} \bigg[ C_j \binom{2\ell-2j}{\ell-j+p}+C_{\ell-j} \binom{2j}{j+p} \bigg] \, ,
\end{align}
where $C_n$ denote the Catalan numbers, defined as $C_n = \frac{1}{n+1} \binom{2n}{n}$. The two terms in \eqref{cleadingN} can be written in a more unified form, using the following property: 
\begin{align}
    \sum_{j=0}^{\ell} \binom{2\ell}{2j} \beta_{[2\ell-2j,2j],[2l]}^{(0)} = 2 \, .
\end{align}
Because of this relation, the first term in \eqref{cleadingN} can be absorbed into the sum over $p$ in the second term, so that we rewrite
\begin{align}\label{cleadingN2}
\begin{split}
     \mathcal{C}^{(0)}_{\cD}(\lambda) &= -2 \sum_{\ell=1}^\infty (-1)^{\ell} (2\ell+1)\,\zeta(2\ell+1)\Big(\frac{\lambda}{16\pi^2}\Big)^\ell \cr
    &\sum_{p=1}^{\ell} \vev{\widehat{\cO}_{[2p]}}_\cD^{(0)} 2^p \sum_{j=0}^{\ell} \binom{2\ell}{2j} \bigg[ C_j \binom{2\ell-2j}{\ell-j+p}+C_{\ell-j} \binom{2j}{j+p} \bigg] \, . 
\end{split}
\end{align}
The sum over $j$ of the above expression can be further simplified using the symmetry of the binomial and Vandermonde's identity:
\begin{equation}
    \begin{split}
        \sum_{j=0}^{\ell} \binom{2\ell}{2j}\, \bigg[ C_j \binom{2\ell-2j}{\ell-j+p}+C_{\ell-j} \binom{2j}{j+p} \bigg] &= 2\sum_{j=0}^{\ell} \binom{2\ell}{2j}\,  \binom{2j}{j+p} C_{\ell-j} \cr
    &= \frac{2}{2\ell+1} \binom{2\ell+1}{\ell+p+1} \binom{2\ell+1}{\ell+p} \, ,
    \end{split}
\end{equation}
and so
\begin{align}
    \mathcal{C}^{(0)}_{\cD}(\lambda) = -4 \sum_{\ell=1}^\infty (-1)^{\ell} \,\zeta(2\ell+1)\Big(\frac{\lambda}{16\pi^2}\Big)^\ell \sum_{p=1}^{\ell}  2^p \binom{2\ell+1}{\ell+p+1} \binom{2\ell+1}{\ell+p} \vev{\widehat{\cO}_{[2p]}}_\cD^{(0)}\,.
\end{align}
Finally, substituting in the result of $\vev{\widehat{\cO}_{[2p]}}_\cD^{(0)}$ given in \eqref{eq:3ptOhat}, we arrive at
\begin{align}
    \cC_{\cD}^{(0)} (\lambda)=-8  \sum_{\ell=1}^\infty(-1)^{\ell+1} \zeta(2\ell+1) \Big(\frac{\lambda}{16\pi^2}\Big)^\ell \sum_{p=1}^{\ell} (-1)^{p} \binom{2\ell+1}{\ell+p+1} \binom{2\ell+1}{\ell+p} \, .
\end{align}
The summation over $p$ in the above expression can be done explicitly, and we find that the integrated correlator in the large-$N$ limit is given by the following remarkably simple formula:
\begin{align}\label{CN_pert_compact}
    \cC^{(0)}_{\cD} (\lambda) =4   \sum_{\ell=1}^{\infty}  (-1)^{\ell+1} {\zeta (2 \ell{+}1)   } \left[\binom{2 \ell+1}{\ell}^2-\binom{2 \ell+1}{\ell}\right] \left( {\lambda \over 16\pi^2}\right)^{\ell} \,.
\end{align}
This all-loop expression of the integrated correlator at leading order in the planar limit is one of our main results. As an example, we write down the first few perturbative orders of the integrated correlator:
\begin{align} \label{eq:CD_pert}
    \cC^{(0)}_{\cD}(\lambda) = \frac{3 \lambda \zeta(3)}{2 \pi ^2}-\frac{45 \lambda^2 \zeta(5)}{32 \pi ^4} +\frac{595 \lambda^3 \zeta(7)}{512 \pi ^6}-\frac{7875 \lambda^4 \zeta(9)}{8192 \pi ^8}  + O(\lambda^{5})\,. 
\end{align}  

\paragraph{Comparison with perturbative results in Feynman integrals.} 
The perturbative expansion for the integrated correlator \eqref{eq:CD_pert}, obtained from the localised matrix model, can be checked against the explicit computation for the un-integrated four-point correlator $\langle \cO_2 \cO_2 \cD \cD \rangle$, which has been computed at one and two loops \cite{Jiang:2019xdz,Jiang:2019zig} and more recently at three loops \cite{Jiang:2023uut}. More specifically, the integral of the (un-integrated) correlator over the measure \eqref{eq:measure} should reproduce \eqref{eq:CD_pert} order by order in perturbation theory.  We perform this analysis in detail in appendix \ref{app:loops}.

We find that our result \eqref{eq:CD_pert} matches exactly the expression given in \cite{Jiang:2019xdz,Jiang:2019zig, Jiang:2023uut} by integrating out the spacetime dependence of the correlator for the first two loops. However, the $O(\lambda^3)$ three-loop result in \eqref{eq:CD_pert} does not agree with the proposed expression given in \cite{Jiang:2023uut} (once again after integrating out the spacetime dependence). We believe the integral basis used in \cite{Jiang:2023uut} for constructing the three-loop integrand for $\langle \cO_2 \cO_2 \cD \cD \rangle$ may not be complete, which is the reason for the discrepancy at three loops. See appendix \ref{app:loops} for a more detailed discussion about the agreement at one and two loops and the mismatch at three loops.

\subsubsection{Subleading-order expression}

We now analyse the subleading contribution in the large-$N$ expression. From \eqref{cDlead+sub}, the integrated correlator at the subleading order can be written as 
\begin{align}\label{cDsub}
            \mathcal{C}^{(1)}_{\cD}(\lambda) &= \, -2\sum_{\ell=1}^\infty (-1)^{\ell} (2\ell+1)\,\zeta(2\ell+1)\Big(\frac{\lambda}{8\pi^2}\Big)^\ell \, \times \\
    &\Bigg[ 2\vev{ \widehat{\cO}_{[2\ell]}}_\cD^{(1)} \, + \sum_{j=0}^{\ell}\binom{2\ell}{2j} \sum_{p=1}^{\ell-1} \beta_{[2\ell-2j,2j],[2p]}^{(0)} \vev{ \widehat{\cO}_{[2p]}}_\cD^{(1)}   + \sum_{j=1}^{\ell-1} \binom{2\ell}{2j} \vev{ \widehat{\cO}_{[2\ell-2j,2j]}}_\cD^{(0)} \cr 
    +\, \frac{1}{2} &\sum_{j=0}^{\ell}\binom{2\ell}{2j} \! \left( \sum_{p=2}^{\ell-1} \sum_{p_1=1}^{p-1} \beta_{[2\ell-2j,2j],[2 p_1, 2p-2p_1]}^{(0)} \vev{ \widehat{\cO}_{[2p_1, 2p-2p_1]}}_\cD^{(0)} \!+\! \sum_{p=1}^{\lfloor \frac{\ell-1}{2} \rfloor}\! \beta_{[2\ell-2j,2j],[2p,2p]}^{(0)} \vev{ \widehat{\cO}_{[2p,2p]}}_\cD^{(0)} \right) \! \Bigg]  \, . \nonumber
\end{align}
Similarly to the leading-$N$ case, using \eqref{eq:beta0}, the first two terms evaluate to
\begin{align}
    2\vev{ \widehat{\cO}_{[2\ell]}}_\cD^{(1)} + \sum_{j=0}^{\ell}\binom{2\ell}{2j} \sum_{p=1}^{\ell-1} \beta_{[2\ell-2j,2j],[2p]}^{(0)} \vev{ \widehat{\cO}_{[2p]}}_\cD^{(1)} = \sum_{p=1}^\ell \frac{2^{p-\ell+1}}{2\ell {+} 1} \binom{2\ell {+} 1}{\ell {+} p{+}1}\binom{2\ell+1}{\ell{+}p} \vev{ \widehat{\cO}_{[2p]}}_\cD^{(1)} \, .
\end{align}
To proceed, we use the expression for $\beta_{[2\ell-2j,2j],[2p_1,2p-2p_1]}^{(0)}$ as given in \eqref{eq:b00}:
\begin{align}
   \begin{split}
        \beta_{[2\ell-2j,2j],[2p_1,2p-2p_1]}^{(0)} &= \, {2^{p-\ell} \over 1+\delta_{p_1,p-p_1}} \bigg[\binom{2j}{j {-} p {+} p_1} \binom{2\ell {-} 2j}{\ell {-} j{-}p_1} +\binom{2j}{j {-} p_1}\binom{2\ell {-} 2j}{\ell {-} j{-}p {+} p_1} \\
    &+\frac{\ell {-} j {-} p}{j {+} 1} \binom{2j}{j} \binom{2 \ell {-} 2j}{\ell {-} j {-} p}+\frac{j {-} p}{\ell {-} j{+}1} \binom{2\ell {-} 2j}{\ell {-} j} \binom{2j}{j{-}p} \bigg] \, .
   \end{split}
\end{align}
We then note that the two terms in the final line of \eqref{cDsub} can be combined to 
\begin{align} \label{eq:lastline}
   \begin{split}
        &\frac{1}{2} \sum_{j=0}^{\ell}\binom{2\ell}{2j} \sum_{p=2}^{\ell-1} \sum_{p_1=1}^{p-1} 2^{p-\ell} \bigg[\binom{2j}{j {-} p {+}p_1} \binom{2\ell{-}2j}{\ell {-} j {-} p_1} +\binom{2j}{j {-}p_1}\binom{2\ell {-} 2j}{\ell {-} j {-} p {+} p_1} \\
    &+\frac{\ell {-} j {-} p}{j {+} 1} \binom{2j}{j} \binom{2\ell {-} 2j}{\ell {-} j {-} p}+\frac{j {-} p}{\ell {-} j {+}1} \binom{2\ell {-} 2j}{\ell {-} j} \binom{2j}{j {-} p} \bigg] \vev{ \widehat{\cO}_{[2p_1, 2p-2p_1]}}_\cD^{(0)} \,. 
   \end{split}
\end{align}
Similarly to the leading-$N$ case, the above binomial sums can be simplified further. In particular, we can perform the $j$ sum first, using the symmetry of the binomial and Vandermonde's identity. Doing so, we find the final line of \eqref{cDsub}, namely \eqref{eq:lastline},  reduces to
\begin{align} \label{eq:final-1}
    \sum_{p=2}^{\ell-1} \sum_{p_1=1}^{p-1} 2^{p-\ell} \binom{2\ell}{\ell {-} p}  \bigg[ \binom{2\ell}{\ell {-} p {+} 2p_1} + \binom{2\ell}{\ell {+} p {+}1} \bigg] \vev{ \widehat{\cO}_{[2p_1, 2p-2p_1]}}_\cD^{(0)} \, . 
\end{align}
Note that if $p=\ell$ in the above equation, then it becomes
\begin{align}
    \sum_{p_1=1}^{\ell-1} \binom{2\ell}{2p_1} \vev{ \widehat{\cO}_{[2p_1, 2l-2p_1]}}_\cD^{(0)}  = \sum_{j=1}^{\ell-1} \binom{2\ell}{2j} \vev{ \widehat{\cO}_{[2\ell-2j, 2j]}}_\cD^{(0)}\, ,
\end{align}
Therefore the $\sum_{j=1}^{\ell-1} \binom{2\ell}{2j} \vev{ \widehat{\cO}_{[2\ell-2j, 2j]}}_\cD^{(0)}$ term in the second line of \eqref{cDsub} can be absorbed into the sum over $p_1$ in the final line as given in \eqref{eq:final-1}. In conclusion, \eqref{cDsub} becomes
\begin{align} \label{cDsub2}
    \mathcal{C}^{(1)}_{\cD}(\lambda) =& \, -2\sum_{\ell=1}^\infty (-1)^{\ell} (2\ell+1)\,\zeta(2\ell+1)\Big(\frac{\lambda}{8\pi^2}\Big)^\ell \,  \times\\
    &\Bigg[ \sum_{p=1}^\ell \frac{2^{p-\ell+1}}{2\ell+1} \binom{2\ell+1}{\ell+p+1}\binom{2\ell+1}{\ell+p} \vev{ \widehat{\cO}_{[2p]}}_\cD^{(1)} \cr
    + & \sum_{p=2}^{\ell} \sum_{p_1=1}^{p-1} 2^{p-\ell} \binom{2\ell}{\ell {-} p}  \bigg(\binom{2\ell}{\ell {-} p {+} 2p_1} + \binom{2\ell}{\ell {+} p{+}1} \bigg) \vev{ \widehat{\cO}_{[2p_1, 2p-2p_1]}}_\cD^{(0)} \Bigg] \, . \nonumber
\end{align}
Using the expression of $\vev{ \widehat{\cO}_{[2p]}}_\cD^{(1)}$ given in \eqref{eq:3ptOhat} and $\vev{ \widehat{\cO}_{[2p_1, 2p-2p_1]}}_\cD^{(0)}$ from \eqref{eq:multi-trace}, we obtain 
\begin{align}\label{cDsub4}
    \mathcal{C}^{(1)}_{\cD}(\lambda) =& \, -4\sum_{\ell=1}^\infty (-1)^{\ell} (2\ell+1)\,\zeta(2\ell+1)\Big(\frac{\lambda}{16\pi^2}\Big)^\ell \, \times \\
    &\Bigg[2 \sum_{p=2}^\ell \binom{2\ell}{\ell-p} (-1)^p \sum_{p_1=1}^{p-1} \binom{2\ell}{\ell {-} p {+} 2p_1} \cr
    &+ \sum_{p=2}^\ell (-1)^p (p-1) \bigg(\frac{3p}{2\ell+1} \binom{2\ell+1}{\ell {+} p {+} 1}\binom{2\ell+1}{\ell+p} + 2 \binom{2\ell}{\ell+p} \binom{2\ell}{\ell+p+1} \bigg)  \Bigg] \, . \nonumber
\end{align}
The first line sums to 
\begin{align}
    2 \sum_{p=2}^\ell (-1)^p \binom{2\ell}{\ell{-}p}  \sum_{p_2=1}^{p-1} \binom{2\ell}{\ell{-}p{+}2p_2} = 2\binom{2\ell{-}1}{\ell}^2-\binom{2l}{l} \, ,
\end{align}
while the second line sums to
\begin{align}
    \sum_{p=2}^\ell (-1)^p (p{-}1) \bigg[\frac{3p}{2\ell{+}1} \binom{2\ell{+}1}{\ell{+}p{+}1}\binom{2\ell+1}{\ell+p} + 2 \binom{2\ell}{\ell{+}p} \binom{2\ell}{\ell{+}p{+}1} \bigg] = \frac{1}{2} \binom{2\ell}{\ell} \left(\binom{2\ell}{\ell{-}1} -\ell\right) \, .
\end{align}
Substituting these into \eqref{cDsub4}, we find the subleading term is given by the following compact expression:
\begin{align} \label{eq:subleading}
    \cC_{\cD}^{(1)}(\lambda) =2 \sum_{\ell=1}^\infty (-1)^{\ell-1} \zeta (2 \ell{+}1) \, (\ell{+}1) \left[ \binom{2 \ell+1}{\ell}^2 -(\ell+2) \binom{2\ell+1}{\ell}\right]\left(\frac{\lambda}{16 \pi^2} \right)^\ell \, ,
\end{align}
which is very similar to the leading order term $\cC^{(0)}_{\cD} (\lambda)$ as given in \eqref{CN_pert_compact}.
We can write down the first few perturbative orders: 
\begin{align}
    \cC_{\cD}^{(1)}(\lambda) = -\frac{45 \lambda^2 \zeta(5)}{32 \pi^4} + \frac{525 \lambda^3 \zeta(7)}{256 \pi^6} -\frac{4725 \lambda^4 \zeta(9)}{2048 \pi^8} + O(\lambda^5) \, .
\end{align}
We note that the $1/N$ correction begins at two loops.  The four-point correlator has only been computed in the strictly planar limit, up to three loops \cite{Jiang:2019xdz,Jiang:2019zig, Jiang:2023uut}, as we reviewed in appendix \ref{app:loops}. Therefore we cannot compare the subleading contribution in \eqref{eq:subleading} with any results in the literature, as we did in the case of the leading planar contribution. However, \eqref{eq:subleading} can represent a useful constraint for any future field theory computation on $\mathbb R^4$ beyond the planar limit.

\subsection{Exact result in $\lambda$ and strong coupling expansion}

It is easy to see that the perturbative series in both the leading large-$N$ limit, namely \eqref{CN_pert_compact}, and its subleading contribution, \eqref{eq:subleading}, are convergent with a finite convergent radius $|\lambda| < \pi^2$, in analogy with many physical quantities in $\mathcal{N}=4$ SYM. We can then resum the perturbative series and obtain a closed form result for $\cC_{\cD} (\lambda)$ at each order in $1/N$ that is \textit{exact for any $\lambda$}. This can be done by using the following integral representation of $\zeta (2 \ell{+}1)$:
\begin{align} \label{eq:id-zeta}
\zeta(2\ell{+}1) = {2^{2\ell} \over \Gamma(2\ell {+} 2)} \int_0^{\infty} dw {w^{2\ell+1} \over \sinh^2(w) }\, .
\end{align}
Therefore the resummation of the perturbative expansion \eqref{CN_pert_compact} for $\cC^{(0)}_{\cD}$ using \eqref{eq:id-zeta} leads to the following exact expression, written in terms of Bessel functions of the first kind $J_{\nu}(x)$:
\begin{align}\label{eq:CD_exact}
     \cC^{(0)}_{\cD} (\lambda) =-  \int_0^{\infty} {8w\, dw \over  \sinh(w)^2 }   \frac{ \left(J_0\left(v \right)-1\right)
   J_1\left(v \right)}{v} \, , 
\end{align}
where $v=w \sqrt{\lambda }/\pi$.  The same analysis applies to the subleading contribution \eqref{eq:subleading}, and we find
\begin{align} \label{eq:CD_exact2}
     \cC^{(1)}_{\cD} (\lambda)  = \int_0^{\infty} {2w \, dw \over \sinh(w)^2}  \left[ J_1\left(v \right) \left( J_1\left(v \right)- {v\over 2} \right)  - \left(J_0\left(v \right)-1\right)^2\right]\, . 
\end{align}
The above integral representations are the analytic continuations of the perturbation expansions given in \eqref{CN_pert_compact} and \eqref{eq:subleading}, and are well-defined for any $\lambda>0$, in particular beyond the convergence radius, \textit{i.e.} $\lambda>\pi^2$.

The exact results allow us to go beyond weakly coupled perturbation theory, so we can study the strong-coupling expansion of the integrated correlator. We express the Bessel function and the product of Bessel functions in terms of Mellin–Barnes type of integrals
\begin{align}
\begin{split}
    J_{\nu}(x) &= {1\over 2\pi \ii} \int_{-\ii \, \infty}^{ \ii\, \infty} {\Gamma(-t)\, x^{\nu+2t} \over 2^{\nu+2t}\, \Gamma(\nu+t+1)} \, , \cr
J_{\mu}(x) J_{\nu}(x) &= {1\over 2\pi \ii} \int_{-\ii \, \infty}^{ \ii\, \infty} {\Gamma(-t)\Gamma(2t+\mu+\nu+1)\, x^{\mu+\nu+2t} \over 2^{\mu+\nu+2t}\, \Gamma(\mu+t+1)\Gamma(\nu+t+1)\Gamma(\mu+\nu+t+1)}\,.
\end{split}
\end{align}
We therefore rewrite the integrated correlator in the following Mellin-Barnes integral:
\begin{align}
     \begin{split}
         \cC^{(0)}_{\cD} (\lambda) &= -{1 \over 2\pi \, \ii} \int_{-\ii \infty}^{\ii \infty}  \int_0^{\infty} {dw \over \sinh^2(w)}  { \lambda^t w^{2 t+1} \Gamma (-t) \over 2^{2 t-2} \, \pi ^{2 t} \Gamma (t+2) }\left( \frac{2^{2 t+1}
   \Gamma \left(t+\frac{3}{2}\right)}{  \pi ^{\frac{1}{2}}  \Gamma
   (t+2)}- 1 \right) \cr 
   &= -{1 \over 2\pi \, \ii} \int_{-\ii \infty}^{\ii \infty}   { \lambda^t   \Gamma (-t)  \zeta (2 t+1) \Gamma (2 t+2) \over 2^{4 t-2} \, \pi ^{2 t} \Gamma (t+2) }\left( \frac{2^{2 t+1}
   \Gamma \left(t+\frac{3}{2}\right)}{  \pi ^{\frac{1}{2}}  \Gamma
   (t+2)}- 1 \right) \, ,
     \end{split}
\end{align}
where in the second line we used the integral identity \eqref{eq:id-zeta}. 
Closing the contour along positive $t$ and picking the residues at $t=1, 2, \ldots $ reproduces the perturbative result as given in \eqref{CN_pert_compact}, whereas closing the contour on the negative $t$ side and picking the residues at $t=0, -1, -3/2, -5/2,  \ldots $ leads to the strong coupling expansion of the integrated correlator. Explicitly, the strong coupling expansion for $\cC_\cD^{(0)}$ takes the following form: 
\begin{align} \label{eq:strong}
    \cC_\cD^{(0)}(\lambda)\vert_{\rm strong} \sim   2 - \frac{4\pi ^2}{3 \lambda } - \sum_{n=1}^{\infty}   \frac{16 n \zeta (2 n+1) \Gamma
   \left(n-\frac{1}{2}\right)^2 \Gamma \left(n+\frac{1}{2}\right)}{  \lambda ^{  n+\frac{1}{2}}\, \pi ^{3/2}\, \Gamma (n)}  \, . 
\end{align}
Proceeding in a similar way for the subleading contribution $\cC_\cD^{(1)}$, we find that its strong coupling expansion is given by
\begin{align} \label{eq:strong2}
    \cC_\cD^{(1)}(\lambda)\vert_{\rm strong} \sim -2 \gamma -2  -\log \left({\lambda \over 16\pi^2 }\right) + \sum_{n=1}^{\infty}  \frac{8 n  \zeta (2 n+1) \Gamma \left(n-\frac{1}{2}\right)
   \Gamma \left(n+\frac{1}{2}\right)^2}{ \lambda ^{  n+\frac{1}{2}}\, \pi ^{3/2} \Gamma (n)}  \, ,
\end{align}
where $\gamma$ is the Euler-Mascheroni constant.

In the holographic dictionary, $\lambda=L^4/\alpha'^2$ (where $\alpha'$ is the square of the string length scale and $L$ is the $AdS_5$ length scale), so the $1/\lambda$ terms correspond to the small-$\alpha'$ corrections in the dual type IIB string theory. Therefore, the leading term (\textit{i.e.} the constant term) is the supergravity contribution to the integrated correlator and the rest corresponds to the stringy corrections. We will discuss the holographic interpretation in more detail in the next section, and we will see that the result agrees with the expectations from type IIB string theory.

\subsubsection{Resurgent analysis of the strong coupling expansion}

It is important to notice that, unlike the small-$\lambda$ expansion \eqref{CN_pert_compact}, the large-$\lambda$ expansion \eqref{eq:strong} is only asymptotic (as is the subleading contribution).  We now analyse  the asymptotic series by the modified Borel transformation following \cite{Arutyunov:2016etw} (see also \cite{Dorigoni:2021guq, Hatsuda:2022enx} in the context of integrated correlators in $\mathcal{N}=4$ SYM). For an asymptotic series, denoted as $f(x)$, we have the modified Borel transform
\begin{align}
    \mathcal{B}: \qquad  f(x) =\sum^{\infty}_{n=1}c_n\, x^{-2n-1} \quad \rightarrow \quad 
   \hat{f}(z) = \sum^{\infty}_{n=1} { c_n \over \zeta(2n{+}1) \Gamma(2n{+}2)} (2z)^{2n+1}   \, , 
\end{align}
and the Borel resummation of $f(x)$ is then given by
\begin{align} \label{eq:borel}
   \mathcal{S}_{\theta} f(x) = {x \over 2} \int_0^{e^{\ii \theta} \infty} {dz \over \sinh^2(x z)} \, \hat{f}(z) \, ,
\end{align}
where we have again used the integral identity \eqref{eq:id-zeta}.  The function $ \mathcal{S}_{\theta} f(x)$ defines an analytic function for $x>0$ and $\theta \in (-\pi/2, \pi/2)$, if the integral \eqref{eq:borel} is well-defined. 

In our case, $x=\sqrt{\lambda}$, and for the leading large-$N$ expansion, as shown in \eqref{eq:strong}, the coefficients $c_n$ are given by
\begin{align}
    c_n = \frac{16\, n\, \zeta (2 n+1) \Gamma
   \left(n-\frac{1}{2}\right)^2 \Gamma \left(n+\frac{1}{2}\right)}{   \pi ^{3/2} \, \Gamma (n)} \, .
\end{align}
The above procedure, in particular the formula \eqref{eq:borel}, leads to the Borel resummation of \eqref{eq:strong}, which takes the following form: 
\begin{align} \label{eq:borel-int}
\mathcal{S}_{\theta} \cC^{(0)}_{\cD} (\lambda)\vert_{\rm strong} =  2 -\frac{4\pi ^2}{3 \lambda }  -\frac{16\sqrt{\lambda}}{3} \int_0^{e^{\ii \theta} \infty} {  z^3 \, dz \over \sinh^2(z \sqrt{\lambda})}\,  _3F_2\left(\frac{1}{2},\frac{1}{2},\frac{3}{2};1,\frac{5}{2};z^2\right)  \, .
\end{align} 
We see, however, that for the asymptotic series \eqref{eq:strong}, the Borel resummed result given above is not well defined, because the hypergeometric function $_3F_2$ has a cut along $[1, \infty]$ and the integral \eqref{eq:borel-int} on the real axis is ill-defined. This implies that the asymptotic series \eqref{eq:strong} is not Borel resummable. The standard resurgence arguments suggest that the large-$\lambda$ expansion \eqref{eq:strong} requires the addition of exponentially small non-perturbative terms, which are obtained as
\begin{align}
 \Delta \cC^{(0)}_{\cD} (\lambda)\vert_{\rm strong} :=&\, (\mathcal{S}_{\theta \rightarrow +0}  - \mathcal{S}_{\theta \rightarrow -0} )\cC^{(0)}_{\cD} (\lambda)\vert_{\rm strong}   
\cr 
=&\, - \frac{16\sqrt{\lambda}}{3} \int_1^{ \infty} {  z^3 \, dz \over \sinh^2(z \sqrt{\lambda})}\, {\rm Disc}\, _3F_2\left(\frac{1}{2},\frac{1}{2},\frac{3}{2};1,\frac{5}{2};z^2\right)\, .
\end{align}
The discontinuity of $_3F_2$ is given by\footnote{The discontinuity can be obtained using the relation $\partial_z \left[ z^3\, _3F_2\left(\frac{1}{2},\frac{1}{2},\frac{3}{2};1,\frac{5}{2};z^2\right)\right] = 3 z^2 \,
   _2F_1\left(\frac{1}{2},\frac{1}{2};1;z^2\right)$ and the known discontinuity of $_2F_1(a,b;c;x)$.}
\begin{align}
{\rm Disc}\, _3F_2\left(\frac{1}{2},\frac{1}{2},\frac{3}{2};1,\frac{5}{2};z^2\right) = \frac{3\, \ii}{\pi ^2} G_{3,3}^{2,3}\left(z^2|
\begin{array}{c}
 -\frac{1}{2},\frac{1}{2},\frac{1}{2} \\
 0,0,-\frac{3}{2} \\
\end{array}
\right)-\frac{3\, \ii\, \pi }{4 z^3}\, ,
\end{align}
where $G_{3,3}^{2,3}$ is the Meijer G-function. To perform the integral, we shift the integration variable $z\rightarrow z+1$ and we expand $\sinh^2(z \sqrt{\lambda})$ as an infinite sum of exponential functions, getting to:
\begin{align}\label{eq:woldsheet_inst0}
\Delta \cC^{(0)}_{\cD} (\lambda)\vert_{\rm strong}  &= -\ii   \frac{ 64\sqrt{\lambda} }{\pi ^2} \sum_{n=1}^{\infty}  n\,  e^{-2 n \sqrt{\lambda } } \int_0^{ \infty} dz \,e^{-2 n \sqrt{\lambda} \, z }   \left[ (z+1)^3 G_{3,3}^{2,3}\left((z+1)^2|
\begin{array}{c}
 -\frac{1}{2},\frac{1}{2},\frac{1}{2} \\
 0,0,-\frac{3}{2} \\
\end{array}
\right)\!-\frac{\pi^3 }{4 } \right] \cr 
&= -\ii    \sum_{n=1}^{\infty}  e^{-2 n \sqrt{\lambda } } \left[\frac{32}{\sqrt{\lambda } \, n} + \frac{24}{\lambda \,  n^2} + \frac{5}{\lambda ^{3/2}\, n^3} -\frac{9}{4 \lambda ^2\, n^4} + \frac{123}{64 \lambda ^{5/2} \, n^5} + \ldots \right] \, ,
\end{align}
where we have expanded the integrand in small $z$ to perform the $z$ integral. 
The expression may also be written in terms of Polylogarithms:
\begin{align}
\Delta \cC^{(0)}_{\cD} (\lambda)\vert_{\rm strong}  &= -\ii  \left[ \frac{32 \, \text{Li}_1\! \left(u\right)}{\sqrt{\lambda } } + \frac{24 \, \text{Li}_2\! \left(u\right)}{\lambda} + \frac{5\, \text{Li}_3\! \left(u\right)}{\lambda ^{3/2}} -\frac{9 \, \text{Li}_4\! \left(u\right)}{4 \lambda ^2} + \frac{123 \, \text{Li}_5\! \left(u\right)}{64 \lambda ^{5/2}} +  \ldots\right] \, , 
\end{align}
where we have denoted $u\! :=e^{-2 \sqrt{\lambda }}$. 

Applying the same analysis to the subleading result $\cC^{(1)}_{\cD}$ given in \eqref{eq:strong2}, we find
\begin{align} \label{eq:strong3}
\Delta \cC_{\cD}^{(1)}(\lambda)\vert_{\rm strong}=\int^{\infty}_1 {\sqrt{\lambda}\, dz \over \sinh^2(z\sqrt{\lambda})} \frac{4 z^3 \, {\rm Disc}\, _3F_2\left(\frac{1}{2},\frac{3}{2},\frac{3}{2};1,\frac{5}{2};z^2\right)}{3} \, , 
   \end{align}
and the discontinuity is given by
\begin{align} \label{eq:Disc_strong2}
{\rm Disc}\, _3F_2\left(\frac{1}{2},\frac{3}{2},\frac{3}{2};1,\frac{5}{2};z^2 \right) = \frac{6\, \ii} {\pi ^2}G_{3,3}^{2,3}\left(z^2\left|
\begin{array}{c}
 -\frac{1}{2},-\frac{1}{2},\frac{1}{2} \\
 0,0,-\frac{3}{2} \\
\end{array}
\right.\right)+\frac{3\, \ii\, \pi }{2\, z^{3}} \, . 
   \end{align}
Performing the $z$-integral explicitly yields the final result, which is given as
\begin{align} \label{eq:Disc_strong3}
\Delta \cC_{\cD}^{(1)}(\lambda)\vert_{\rm strong}=\ii \left[ 16\text{Li}_0(u)+{4 \text{Li}_1(u) \over \sqrt{\lambda}}+{5 \text{Li}_2(u) \over 2{\lambda}}+{ \text{Li}_3(u) \over 8 \lambda^{3/2}} - { 21\text{Li}_4(u) \over 128 \lambda^{2}}   + \ldots \right] \,.
   \end{align}

Let us briefly comment on the results \eqref{eq:woldsheet_inst0} and \eqref{eq:Disc_strong3}. We can interpret these formulas as perturbative expansions around different ‘worldsheet instanton’ configurations. Indeed, when translated into the holographic dual string theory language, each exponential term in \eqref{eq:woldsheet_inst0} is mapped as $e^{-2 \sqrt{\lambda} } = e^{-2 L^2/\alpha'}$. In the holographic picture, such exponential terms act as instanton weights for the worldsheet instanton contributions and the perturbative expansion around each exponential term corresponds to fluctuations around a worldsheet classical solution.  Furthermore, a similar resurgent analysis has been performed in \cite{Dorigoni:2021guq, Hatsuda:2022enx}  for the integrated correlator $\langle \cO_2 \cO_2 \cO_2 \cO_2 \rangle$. In the case of $\langle \cO_2 \cO_2 \cO_2 \cO_2 \rangle$, it was found in \cite{Dorigoni:2022cua} that the modular function corresponding to the `instanton factor' $e^{-2 n \sqrt{\lambda} }$, after summing over its $SL(2,\mathbb{Z})$ images, can be written in terms of modular-invariant functions and may be interpreted as the worldsheet instantons of $(p,q)$-strings.  We expect the same interpretation to hold for both $\Delta \cC^{(0)}_{\cD} (\lambda)\vert_{\rm strong}$ and $\Delta \cC^{(1)}_{\cD} (\lambda)\vert_{\rm strong}$. 

\section{$SL(2, \mathbb{Z})$ completion}
\label{sec:sl2z}

In this section, we discuss the $SL(2,\mathbb{Z})$ properties of the determinant integrated correlator $\cC_\cD$.  Correlation functions of superconformal primary operators in $\mathcal{N}=4$ SYM with $SU(N)$ gauge group are $SL(2, \mathbb{Z})$ invariant \cite{Montonen:1977sn,Collier:2022emf,Paul:2022piq, Paul:2023rka,Brown:2023cpz,Brown:2023why}, because these operators are neutral under the bonus $U(1)_Y$ symmetry \cite{Intriligator:1998ig, Intriligator:1999ff}.\footnote{Correlation functions of operators that  are charged under the bonus $U(1)_Y$ symmetry (for example chiral Lagrangians) should in general transform non-trivially under $SL(2, \mathbb{Z})$ transformation. See \cite{Green:2020eyj, Dorigoni:2021rdo} for the study of integrated correlators of this type.} Therefore, $\cC_\cD(\tau, \bar{\tau}; N)$ is also $SL(2, \mathbb{Z})$ invariant, as it can be viewed as a linear combination of correlators of superconformal primary operators. To achieve this, we work in a different regime compared to the previous sections, by taking $N$ large while keeping $\tau$ fixed (instead of the 't Hooft limit), for which the instanton contributions are important.
We propose the $SL(2,\mathbb{Z})$ completions for the first few terms from the strong coupling expansions in the zero instanton sector \eqref{eq:strong} and \eqref{eq:strong2}, by promoting the perturbative terms to the appropriate modular functions. This proposal can be checked against explicit instanton calculations from the matrix model. The analytic evaluation of instanton contributions in the large-$N$ fixed-$\tau$ limit turns out to be complicated, so we compute the contributions for many different values of $N$, and extrapolate the result numerically to large $N$. We find that the numerical result matches the expectations from $SL(2,\mathbb{Z})$ completion of our results. We further compare our $SL(2,\mathbb{Z})$-invariant expression with some known results for string amplitudes in flat space, where the large-$N$ expansion (with fixed Yang-Mills coupling) of the correlator is mapped to the small-$\alpha'$ expansion (with fixed string coupling) in string theory. 

\subsection{Modular invariance}
%Therefore, if we consider the large-$N$ of $\langle  \mathcal{O}_2\mathcal{O}_2 \mathcal{D}  \mathcal{D}  \rangle$ with fixed Yang-Mills coupling $\tau$ (instead of the usual 't Hooft limit considered in the previous section), the coefficients at each order of $1/N$ expansion should be given by modular functions of $\tau, \bar{\tau}$. Indeed, from previous results for integrated correlators for trace operators $\langle  \mathcal{O}_2\mathcal{O}_2 \mathcal{O}_p \mathcal{O}_p \rangle$ \cite{Chester:2019jas,  Paul:2022piq, Dorigoni:2022cua, Paul:2023rka, Brown:2023cpz,Brown:2023why}, we expect the same class of modular functions to appear in the large-$N$ fixed-$\tau$ expansion of the integrated correlator, namely \textit{non-holomorphic Eisenstein series}. 

To infer the modular properties of the integrated correlator $\cC_{\cD}(\tau, \bar \tau; N)$,  we rewrite the strong coupling expansion \eqref{eq:strong} by converting $\lambda$ into the (complexified) Yang-Mills coupling $\tau:=\tau_1 +\ii\, \tau_2$ as given in \eqref{eq:tau_def}, using the relation $\lambda=4\pi N/\tau_2$. We then combine  $N\! \cdot \cC^{(0)}_{\cD}(\lambda)\vert_{\rm strong}$ from \eqref{eq:strong} with $\cC^{(1)}_{\cD}(\lambda)\vert_{\rm strong}$ given in \eqref{eq:strong2}  and get
\begin{align} \label{eq:fixed-tau}
\cC_{\cD}(\tau, \bar \tau; N) =&\, N \left(2 - \frac{\pi \tau_2}{3 N} -{1\over N^{3/2}} \frac{ \zeta(3) \tau^{3/2}_2}{\pi^{3/2}} +  O(N^{-5/2}) \right) -\log(N) \cr
&+ \left(-2 \gamma -2  +\log \left(4\pi \right) +\log \left(\tau_2 \right) +  O(N^{-3/2}) \right) +\ldots \, . 
\end{align}
A few comments about this expression are in order.
Firstly,  we need to be careful when translating $\lambda$ to $\tau_2$, since different orders of the 't Hooft $1/N$ expansion can mix with each other in the large-$N$ fixed-$\tau$ expansion. For example, we see already that the term of order $O(N^{0})$ in \eqref{eq:fixed-tau} receives contributions from both $N \cdot \cC_\cD^{(0)}(\lambda)\vert_{\rm strong}$ and $\cC_\cD^{(1)}(\lambda)\vert_{\rm strong}$. As we will see shortly, these two terms in fact must combine together to form a modular function. Moreover, we have only kept the first few orders at large $N$, because, as we will see, these are the terms that have been computed in the holographic dual picture in string theory in flat space, where the correlator is dual to two gravitons scattering off a D3-brane travelling along the geodesic. We can therefore make an explicit comparison with the string-theory results.
Finally, let us also comment on the $\log(N)$ term, which arises when translating $\lambda$ to $\tau_2$. It should correspond to the logarithmic contribution in small-$\alpha'$ expansion of the annulus amplitude, which accompanies the $\log(\tau_2)$ term.\footnote{A similar term appears in the context of integrated correlators of $\langle \cO_2 \cO_2 (\cO_2)^p (\cO_2)^p  \rangle$ and their generalisations as considered in \cite{Brown:2023why, Paul:2023rka} (see also \cite{Caetano:2023zwe}), where one also finds a universal $\log(p)$ term in the large-charge ({\it i.e.} large-$p$) expansion.}  

In the rest of the section we concentrate on the $\tau_2$-dependent terms, which should be completed to $SL(2, \mathbb{Z})$-invariant modular functions.  
To do so, as we discussed in the above, we reorganise the large-$N$ expansion of $\cC_{\cD}(\tau, \bar \tau; N)$ in \eqref{eq:fixed-tau} as follows
\begin{align} \label{eq:fixed-tau2}
\cC_{\cD}(\tau, \bar \tau; N) =&\, N \left[2 -{1\over N} \left( 2 \gamma +2  -\log \left(4\pi \right) \right) \right] - \log(N) \cr
& - \left[ \frac{\pi \tau_2}{3} -\log \left(\tau_2 \right) \right] -{1\over N^{1/2}} \frac{ \zeta(3) \tau^{3/2}_2}{\pi^{3/2}}  +  O(N^{-3/2}) \, , 
\end{align}
and we now analyse the modular property of each term. The $\tau$-independent constants are of course already modular invariant, and holographically they are dual to the supergravity contributions (corresponding to the tree level and loop contributions). All the other $\tau$-dependent terms, corresponding to stringy corrections, should be completed by modular functions. For any given $N$, the determinant integrated correlator $\cC_{\cD}(\tau, \bar \tau; N)$ can in principle be expressed as a linear combination of integrated correlators of trace operators $\langle  \mathcal{O}_2\mathcal{O}_2 \mathcal{O}_p \mathcal{O}_p \rangle$. Therefore, we expect the same class of modular functions that are relevant for the integrated correlators $\langle  \mathcal{O}_2\mathcal{O}_2 \mathcal{O}_p \mathcal{O}_p \rangle$ to appear in the large-$N$ fixed-$\tau$ expansion of  $\cC_{\cD}(\tau, \bar \tau; N)$. From the previous results for integrated correlators  $\langle  \mathcal{O}_2\mathcal{O}_2 \mathcal{O}_p \mathcal{O}_p \rangle$ with the same integration measure \eqref{eq:measure} \cite{Chester:2019jas,Paul:2022piq,Dorigoni:2022cua, Paul:2023rka,Brown:2023cpz,Brown:2023why},  we know these modular functions are the \textit{non-holomorphic Eisenstein series}, which are defined in terms of lattice sums as follows:
\begin{align} \label{eq;Eisenstein}
E(s; \tau, \bar{\tau}) = \sum_{(m,n)\neq (0,0)} {1\over \pi^s} {\tau^s_2 \over |m+n\tau|^{2s}} =  \sum_{k \in \mathbb{Z}} \mathcal{F}_k (s; \tau_2) e^{2\pi \ii \, k \tau_1} \, . 
\end{align}
The second equality corresponds to the Fourier mode expansion. In particular, the zero Fourier-mode coefficient is given by
\begin{align}
\cF_0(s;\tau_2)= \frac {2\zeta(2s)}{\pi^s} \tau_2^s  +   \frac{2\sqrt \pi \,\Gamma(s-{1\over 2}  ) \zeta(2s-1)}{\pi^s \Gamma(s)}\, \tau_2^{1-s}\,.
\label{eisenzero}
\end{align}
This should be compared with the zero-instanton expansion for the integrated correlator. Analogously, the $k$-th Fourier mode, corresponding to the $k$-instanton sector, is proportional to a $K$-Bessel function:
\begin{align}
\cF_k(s;\tau_2)  = \frac{4}{\Gamma(s)}\,  |k|^{s-{1\over 2}} \, \sigma_{1-2s}(|k|)
\sqrt{\tau_2}\,K_{s- {1\over 2} }(2\pi |k|\tau_2) \,, \qquad {\rm with} \quad k\neq 0\,,
\label{nonzeroeisen}
\end{align}
where $\sigma_{q}(k) :=\sum_{d>0, d|k} d^q$ is the sum of the divisor of $k$. 
A special case arises for $s=1$, which will be relevant to our discussion. Indeed, the definition of $E(1; \tau, \bar{\tau})$ given in \eqref{eq;Eisenstein} requires a regularisation by subtracting a logarithmic divergent $\tau$-independent constant. Explicitly, the Fourier-mode expansion of $E(1; \tau, \bar{\tau})$ is given by
\begin{align} \label{eq:E1}
E(1; \tau, \bar{\tau}) = \sideset{}{'}\sum_{(m,n)\neq(0,0)}  {1\over \pi} {\tau_2 \over |m+n\tau|^2} = {\pi \tau_2 \over 3} - \log(\tau_2) + 2 \sum_{k=1}^{\infty} \sum_{m|k} {1\over m} \left( e^{2\pi i k \tau} + e^{-2\pi i k \tau} \right)\, ,
\end{align}
where the “primed sum” indicates the regularisation. 

Now, if we assume the $SL(2, \mathbb{Z})$ completion of $\cC_{\cD}(\tau, \bar \tau; N)$ is indeed given by the non-holomorphic Eisenstein series as in the case of $\langle  \mathcal{O}_2\mathcal{O}_2 \mathcal{O}_p \mathcal{O}_p \rangle$, we can infer the following $SL(2,\mathbb{Z})$ completion rule to be applied to the zero-instanton sector computed via the matrix model by inverting the Fourier zero-mode \eqref{eisenzero}: 
\begin{align} \label{eq:SL2rule}
\frac{2\zeta(2s)\, \tau^{s}_2 }{\pi^{s}} \rightarrow   E(s; \tau, \bar{\tau})   \, .
\end{align}
Applying this on our result \eqref{eq:fixed-tau} we get:
\begin{align} \label{eq:E1E32}
 \frac{\pi \tau_2}{3}  \rightarrow  E(1; \tau, \bar{\tau}) \,  , \qquad  \frac{\zeta(3)\, \tau^{3/2}_2 }{\pi^{3/2}} \rightarrow  {E(3/2; \tau, \bar{\tau}) \over 2} \,  , 
 \end{align}
so that the proposed $SL(2,\mathbb{Z})$ completion of the integrated correlator reads:
\begin{align} \label{eq:fixed-tauSL2Z}
\cC_{\cD}(\tau, \bar \tau; N)= N\! \left[2 -{1\over N} \left( 2 \gamma +2  -\log \left(4\pi \right) \right) \right]\! - \log(N) 
- E(1; \tau, \bar{\tau}) - {E(3/2; \tau, \bar{\tau})\over 2\, N^{1/2}}  +  O(N^{-3/2})   \, . 
\end{align}
It is worth emphasising that in the case of $E(1; \tau, \bar{\tau})$, all the terms in the zero instanton sector (namely the $\tau_2$ term and $\log(\tau_2)$ term) match non-trivially the perturbative results from the matrix model computation, one arising from $N\cdot \cC^{(0)}(\lambda)\vert_{\rm strong}$ and the other from $\cC^{(1)}(\lambda)\vert_{\rm strong}$. This is clearly a non-trivial confirmation of the proposal. Furthermore, from \eqref{eisenzero} we see that, besides the term $\zeta(3)\tau_2^{3/2}$,  the zero mode of $E(3/2; \tau, \bar{\tau})$ also contains a term proportional to $\tau_2^{-1/2}$. Such a term must arise from the NNLO contribution in the planar expansion, which we have not computed in this paper. More concretely, the NNLO integrated correlator $N^{-1} \cdot \cC_\cD^{(2)}(\lambda)\vert_{\rm strong}$ must contain $N^{-1} \cdot \lambda^{1/2} \sim N^{-1/2} \tau_{2}^{-1/2}$ with a precise coefficient $-{2 \zeta(2)}/{\pi^{3/2}}$, so that it can be combined with the $N^{-1/2}\zeta(3)\tau_2^{3/2}$ term to form $E(3/2; \tau, \bar{\tau})$. 

Hence we see that the $SL(2,\mathbb{Z})$ completion is rather powerful, allowing us to predict NNLO contributions in the 't Hooft expansion. The predictive power from $SL(2,\mathbb{Z})$ completion goes beyond the zero-instanton sector. Indeed, considering the higher Fourier coefficients for non-holomorphic Eisenstein series as given in \eqref{eq:E1} for $E(1; \tau, \bar{\tau})$ and \eqref{nonzeroeisen} for $E(s; \tau, \bar{\tau})$ with $s>1$,  they predict the instanton contributions to the integrated correlator $\cC_{\cD}(\tau, \bar{\tau}; N)$.  
In particular, the expectation for the one-instanton term for the first few orders in the $1/N$ expansion is
 \begin{align}  \label{eq:leadingSL}
\cC_{\cD; \, 1\textrm{-inst}}(\tau, \bar{\tau}; N)=e^{2\, \ii \pi \tau}  \left[ -2 - {1\over N^{1/2} } \left(\frac{2}{\sqrt{\pi }} + O(\tau_2^{-1})  \right)  + \ldots \right] \, , 
\end{align} 
where the leading large-$N$ term comes from $E(1;\tau, \bar{\tau})$ and the $N^{-1/2}$ term is from $E(3/2;\tau, \bar{\tau})$. 

This prediction can be checked against explicit instanton computations from the matrix model, as we will discuss in the next subsection.

\subsection{Match with instanton computations in the matrix model}

To confirm the proposed $SL(2,\mathbb{Z})$ completion of our result, we will now check the predicted non-perturbative instanton contributions explicitly. It turns out that it is enough to consider only the one-instanton contribution. As argued in \cite{Collier:2022emf}, due to $SL(2, \mathbb{Z})$-invariance the multi-instanton contributions are in fact determined by the zero- and one-instanton terms. Therefore, the goal is to confirm the one-instanton expression as given in \eqref{eq:leadingSL}. From a technical point of view, we will treat the instanton contributions similarly to the zero-instanton perturbative analysis from sections \ref{sec:3} and \ref{sec:4}. More specifically, we show that the instanton contributions can be decomposed into infinite sums of normalised three-point and two-point functions of normal ordered operators in the presence of determinant operators. The procedure that leads to this rewriting is very similar to section \ref{sec:4}, so in this section we will just outline the main steps.

We return back to the localised partition function \eqref{eq:Z_matrixModel_tot} to include the one-instanton sector.  The one-instanton contribution to the integrated correlator comes from the following term from the partition function \footnote{Please note the difference between the calligraphic $\cZ_{\rm inst}^{(1)}(\tau, \tau'_p; m)$, representing the whole one-instanton contribution to the partition function (already integrated over the eigenvalues), and the integrand $Z^{(1)}_{\rm inst}(\tau'_p, a;m)$, corresponding to the one-instanton Nekrasov partition function.}:
\begin{align}
\mathcal{Z}_{1\textrm{-inst}} (\tau, \tau'; m) = e^{2\pi \ii \tau} \int  d\mu(a_i) \left\vert Z_{\rm classic}(\tau, \tau', a)\,  Z_{1\textrm{-inst}}(\tau', a;m) \right\vert^2\phantom{\bigg|} \, , 
\end{align}
where the measure $d\mu(a_i)$ is given in \eqref{eq:measurea}, and
\begin{align}
 Z_{\rm classic}(\tau, \tau', a)=     \exp\bigg(\ii\, \pi \tau \sum_i a_i^2 +\ii \sum_{p>2} \pi^{p/2} \tau'_p \sum_{i} a_i^p \bigg) \, ,  
\end{align}
and the one-instanton partition function $Z_{1\textrm{-inst}} (\tau', a; m)$ is given by \cite{Fucito:2015ofa}
\begin{align}
 Z_{1\textrm{-inst}}(\tau', a; m)=  -m^2 \sum_{\ell=1}^N \exp\! \left[-\ii \sum_{p>2}  \pi^{p/2} \tau_p' \left( {a}_{\ell}^p + (a_{\ell}+2 \ii)^p -2(a_{\ell}+\ii)^p \right) \right]\! \prod_{j \neq \ell } \frac{(a_{\ell j}+\ii)^2-m^2}{(a_{\ell j}+\ii)^2+1} \, .
\end{align}
From the definition \eqref{eq:C_D_def}, the one-instanton contribution to the integrated correlator $\langle \cO_2 \cO_2 \cD \cD \rangle$  can be formally written as
\begin{align}\label{eq:CD_1-inst}
\cC_{\cD; \, 1\textrm{-inst}}(\tau, \bar{\tau}; N)  = \frac{\partial_{\mathcal{D}} \partial_{\bar{\mathcal{D}}} \partial_m^2 \mathcal{Z}_{1\textrm{-inst}}(\tau, \tau'; m) \, {\vert}_{m=0, \tau'=0}}{\partial_{\mathcal{D}} \partial_{\bar{\mathcal{D}}}  \mathcal{Z}(\tau, \tau'; m)\, {\vert}_{m=0, \tau'=0}} - \frac{ \partial_m^2 \mathcal{Z}_{1\textrm{-inst}} (\tau, \tau'; m) \, {\vert}_{m=0, \tau'=0}}{ \mathcal{Z}(\tau, \tau'; m)\, {\vert}_{m=0, \tau'=0}} \, .
\end{align}
As we did for the 1-loop determinant $Z_{1\textrm{-loop}}$ in section \ref{sec:4}, we can rewrite $Z_{1\textrm{-inst}}$ as a sum over multitrace normal ordered operators, schematically, \begin{align}\label{eq:dmZinst}
    \partial_m^2 Z_{1\textrm{-inst}}( a; m)\, {\vert}_{m=0} = \sum_{\ell =0}^{\infty} (2\pi \tau_2)^{-\ell/2} \sum_{\vec r \,\vdash r \leq \ell} c_{\ell;\, \vec r} \, \cO_{\vec r} \, , 
\end{align}
where the coefficients $c_{\ell;\, \vec r}$ are not $\tau_2$-dependent. As discussed in section \ref{sec:3.3}, the partial derivative $\partial_{\mathcal{D}}$ defines the insertion of a determinant operator, and can be expressed as a linear combination of $\partial_{\tau'_p}$. 
When both $\partial_{\mathcal{D}} $ and  $\partial_{\bar{\mathcal{D}}}$ act on $ Z_{\rm classic}(\tau, \tau', a)$, one gets a sum of three-point functions: 
\begin{align}\label{eq:inst_3pt}
  \sum_{\ell =1}^{\infty} (2\pi \tau_2)^{-\ell}   \sum_{\vec r \,\vdash r \leq 2\ell}  c_{2\ell;\, {\vec r}}  \frac{ \langle  \cD\, \cD \, \cO_{\vec r}  \rangle_0 }{\langle  \cD\, \cD  \rangle_0} \, , 
\end{align}
where $\langle \,\, \rangle_0$ is defined according to \eqref{eq:f_Gaussian}. Similarly to the weak coupling case \eqref{eq:CD_as_vevM2}, the sum is constrained to $r>0$ because the contribution from the identity operator (\textit{i.e.} the constant term) cancels with the second term in \eqref{eq:CD_1-inst}. Furthermore, it is clear that only operators with even dimensions can contribute, so that $r$ and $\ell$ are even numbers.  

Compared to the zero-instanton case, here we also encounter contributions from two-point functions of a determinant operator and a trace operator. When $\partial_{\cD}$ acts on $Z_{1\textrm{-inst}}( a, \tau'; m)$ and $\partial_{\bar \cD}$ acts on $ Z_{\rm classic}(\tau, \tau', a)$, we can expand as in \eqref{eq:dmZinst}:
\begin{align}\label{eq:dDdmZinst}
\partial_{\cD}   \partial_m^2  Z_{1\textrm{-inst}}(\tau', a; m)\, {\vert}_{m=0, \tau'=0} = \sum_{\ell =0}^{\infty} (2\pi \tau_2)^{-\ell/2} \sum_{\vec r \,\vdash r \leq \ell}  d_{\ell; \, \vec r}  \,  \cO_{\vec r}   \, ,
\end{align}
where again the coefficients $d_{\ell; \, \vec r}$ are independent of $\tau$. Therefore, we obtain the second contribution to \eqref{eq:CD_1-inst} in terms of two-point functions:
\begin{align} \label{eq:inst-2}
\sum_{\ell =0}^{\infty} (2\pi \tau_2)^{N-\ell \over 2}  \sum_{\vec r \,\vdash r \leq \ell}  d_{\ell;\, \vec r}  \frac{ \langle  \cD \,  \cO_{\vec r}  \rangle_0 }{\langle  \cD\, \cD  \rangle_0 }    \, ,
\end{align}
where the additional factor of $(2\pi \tau_2)^{N/2}$ is due to the fact that there is an additional $\cD$ operator in the denominator, which scales as $(2\pi \tau_2)^{N/2}$ when we rescale $a_i \rightarrow a_i/\sqrt{2\pi \tau_2}$.  Due to the properties of normal ordered operators and the orthogonality of two-point functions, for any given $N$ the above expression \eqref{eq:inst-2} only receives contributions from operators $\cO_{\vec r}$ with $r =N$. This implies that the summation range is restricted to $\ell \geq N$.  
Putting all these together, in summary the one-instanton contribution to the integrated correlator can be expressed as infinite sums of three-point and two-point functions in the Gaussian matrix model:  
\begin{align} \label{eq:1-inst2}
\cC_{\cD; \, 1\textrm{-inst}}(\tau, \bar{\tau}; N)  = e^{2\pi \ii \tau}\! \left[   \sum_{\ell =1}^{\infty} (2\pi \tau_2)^{-\ell} \!  \sum_{\vec r \,\vdash r \leq 2\ell}  c_{2\ell; \, \vec r}  \frac{ \langle  \cD\, \cD \, \cO_{\vec r}  \rangle_0 }{\langle  \cD\, \cD  \rangle_0}  + \sum_{\ell =N}^{\infty} (2\pi \tau_2)^{N-\ell \over 2} \! \sum_{\vec r \,\vdash N}  d_{\ell; \, \vec r} \frac{ \langle  \cD \,  \cO_{\vec r}  \rangle_0 }{\langle  \cD\, \cD  \rangle_0 }  \right] \, , 
\end{align}
where again the first term only receives contributions from operators $\cO_{\vec r}$ with even $r$.  
As expected, we note that \eqref{eq:1-inst2} representing the perturbative expansion around the one-instanton background starts with $(\tau_2)^0$.

\begin{figure}[t!]
  \centering
\includegraphics[keepaspectratio, width=0.75\textwidth]{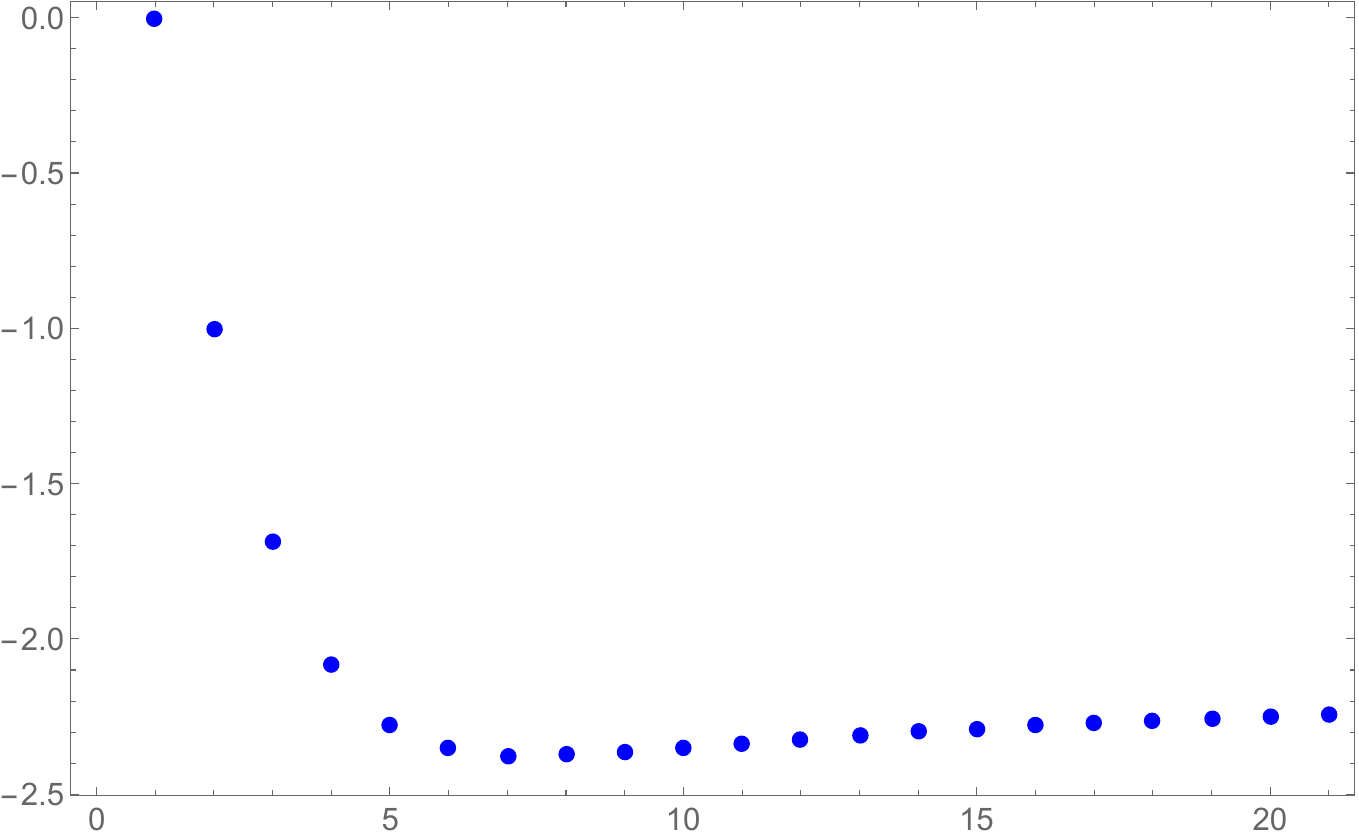}
  \caption{The one-instanton results at the order $O(\tau_2^0)$ computed from the matrix model, up to $N=21$.}
  \label{fig:1-instanton}
\end{figure}

We can now apply the matrix model techniques reviewed in appendix \ref{app:recursion} to evaluate the Gaussian integrals that appear in \eqref{eq:1-inst2}, and obtain the one-instanton contribution to the integrated correlator, which then can be compared against the expected result \eqref{eq:leadingSL} from the $SL(2,\mathbb{Z})$ completion. In particular, focusing on the $\tau_2$-independent contributions (\textit{i.e.} $-2 - {1\over N^{1/2} } \frac{2}{\sqrt{\pi }}$ in \eqref{eq:leadingSL}), they can only arise from the second term in \eqref{eq:1-inst2}. 
Furthermore, we can exploit the relation between the determinant operators and SPOs, as described in appendix \ref{app:2}. In particular, in the large-$N$ limit the determinant operator can be identified with the SPO with dimension $N$ and the difference is irrelevant for the order we consider here. Since any SPO is orthogonal to all multi-trace operators, see \eqref{eq:SPO_multitr_a}, only the single-trace operators contribute to \eqref{eq:inst-2}. This further simplifies the one-instanton calculations. 
However, even with all these simplifications, an analytic analysis of the instanton contribution in the large-$N$ limit (with fixed $\tau$) is still rather difficult. Therefore, we follow a numerical approach. We compute the one-instanton contribution at fixed values of $N$, up to $N=21$, with the explicit data as shown in Fig. \ref{fig:1-instanton}. As one can see that the data clearly shows the asymptotic behaviour approaching $-2$ from below, in agreement with $-2 - {1\over N^{1/2} } \frac{2}{\sqrt{\pi }}  \approx -2.24623$ for $N=21$. We can also extrapolate the large-$N$ properties by fitting the finite-$N$ data with appropriate $1/N^{1/2}$ Taylor series. Doing so, we once again find that the numerical analysis after the large-$N$ extrapolation indeed agrees with the one-instanton result \eqref{eq:leadingSL} as expected from the proposed $SL(2, \mathbb{Z})$ completion.

\subsection{Holographic interpretation from string theory}
Finally, we comment on the comparison between the result \eqref{eq:fixed-tau} and its $SL(2, \mathbb{Z})$ completion \eqref{eq:fixed-tauSL2Z} with the holographic string dual picture of the correlator.

\begin{figure}[t!]
  \centering
\includegraphics[keepaspectratio, width=0.75\textwidth]{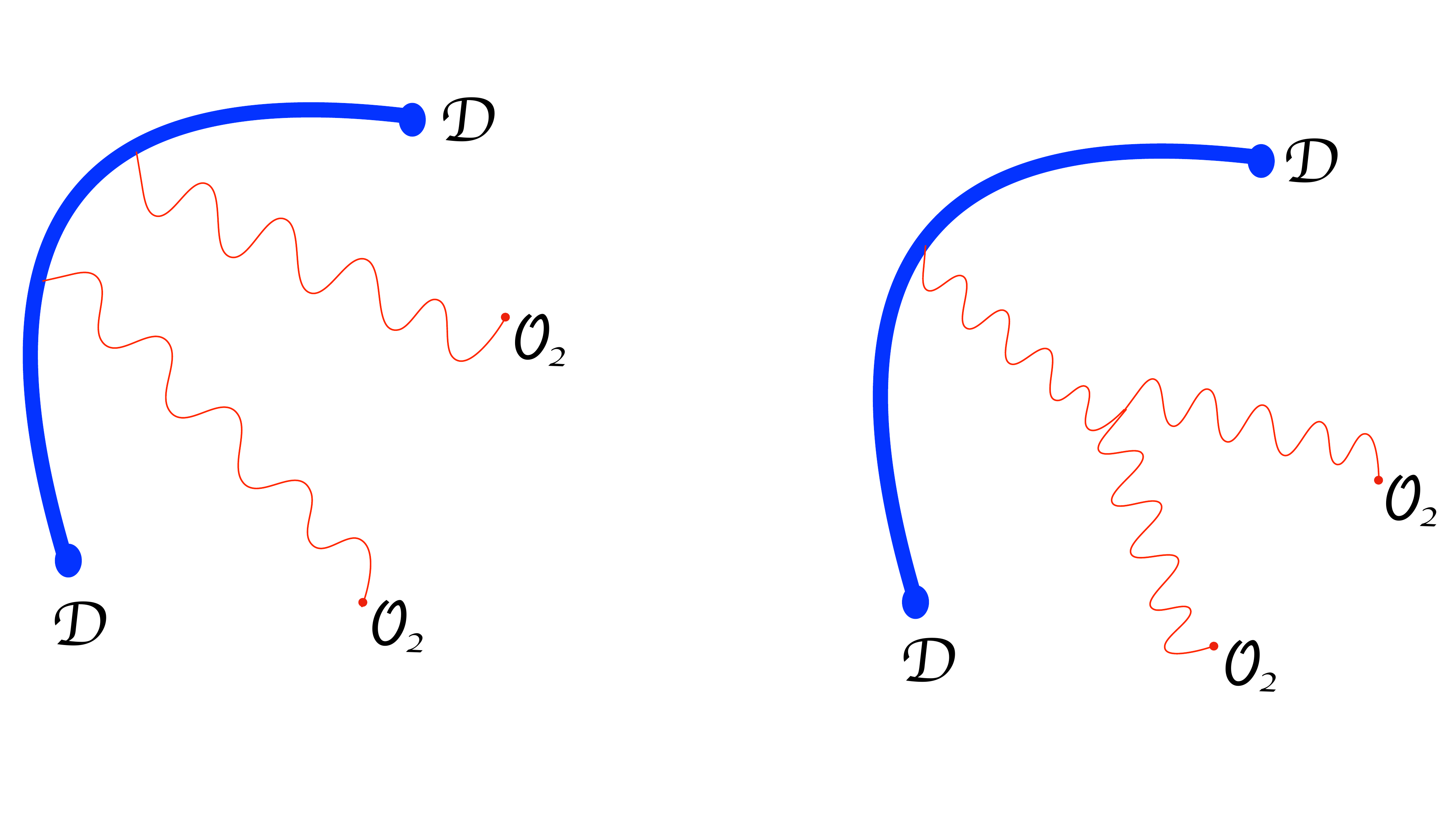}
  \caption{The holographic description of the four-point correlation function of two determinant operators and two half-BPS operators with dimension two. The determinant operators in the large-$N$ limit create a D3-brane in the bulk that travels along the geodesics in AdS, and half-BPS operators represented by the red wavy line are gravitons, which scatter off the D3-brane.}
  \label{fig:gravitonscattering}
\end{figure}

The determinant operators are known to be dual to D3-branes wrapping an $S^3$ inside the $S^5$ in the holographic dual type IIB string theory in $AdS_5 \times S^5$. Therefore, the four-point function $\langle \cO_2 \cO_2 \cD \cD \rangle$ that we study in this paper is dual to the physical process of two gravitons scattering off a D3-brane that moves along the geodesic. This scattering process is schematically shown in Fig. \ref{fig:gravitonscattering}. Importantly, D3-branes are self-dual under $SL(2, \mathbb{Z})$ transformation and gravitons are also neutral under $SL(2, \mathbb{Z})$, therefore this scattering process is modular invariant. This is consistent with the $SL(2, \mathbb{Z})$-invariance of the correlator from the point of view of $\mathcal{N}=4$ SYM. 

Recall the AdS/CFT dictionary 
$\tau  = \tau_s$ and $\lambda = {L^4 \over \alpha'^2 },$ where $\tau_s$ is the string coupling and $L$ is the AdS radius. This implies that the large-$N$ fixed-$\tau$ expansion of the correlator is mapped to the small-$\alpha'$ fixed-$\tau_s$ expansion of string amplitudes.
The disc and annulus contributions to the Brane-Bulk amplitude, which are holographically dual to the planar limit and next-to-planar limit of the four-point correlator we consider here, have been computed in the flat space in \cite{Garousi:1996ad, Hashimoto:1996bf} and \cite{Pasquinucci:1997di, Lee:1997gwa}, respectively. More importantly, the $SL(2,\mathbb{Z})$ completion of this Brane-Bulk amplitude, at least for the first few orders in the $\alpha'$ expansion, is also known. It has been shown that the leading stringy correction \cite{Bachas:1999um, Green:2000ke} as well as the subleading correction \cite{Basu:2008gt, Lin:2015ixa} to this scattering process can be completely determined using supersymmetry and $SL(2, \mathbb{Z})$ invariance. The result can be conveniently summarised as higher derivative terms that contribute to the effective action of the D3-brane. Schematically it reads:
\begin{align}
\int d^{4}x \sqrt{g} \left( \pi E(1; \tau_s, \bar{\tau}_s)\, \alpha'^2 R^2 + \pi^{3/2} E(3/2; \tau_s, \bar{\tau}_s)\, \alpha'^3 D^2 R^2 + {\rm higher~derivatives} \right) \, ,
\end{align}
where we have only kept the derivative corrections of the type $D^{2k} R^2$ that are relevant for the scattering process dual to the four-point correlator we consider here. 
We see that these string corrections are indeed in perfect agreement with the large-$N$ fixed-$\tau$ expansion of the integrated correlator $\cC_{\cD}(\tau, \bar{\tau})$ we discussed earlier. In particular, $\alpha'^2 R^2$  is mapped to the first $1/N$ correction   (\textit{i.e.} the third term) in \eqref{eq:fixed-tauSL2Z} and $\alpha'^3 D^2 R^2$ is mapped to the second $1/N$ correction (\textit{i.e.} the fourth term) in \eqref{eq:fixed-tauSL2Z}. More explicitly, the $\tau_2$ term in $E(1; \tau, \bar{\tau})$ corresponds to the disc amplitude contribution, whereas the $\log(\tau_2)$ term is associated with the annulus amplitude, and the instanton contributions are due to the D(-1)-instanton effects. Similar interpretations can be given to the terms in $E(3/2; \tau, \bar{\tau})$.

 \section{Conclusion and discussion}
 \label{sec:7}
In this paper, we provided a large set of results for the integrated correlator of two chiral primary operators with dimension two in the presence of two determinant operators using supersymmetric localisation. Starting from a weak coupling approach from the matrix model in the large-$N$ and fixed-$\lambda$ regime in the planar limit and beyond, we developed new techniques to compute matrix model observables in terms of Gaussian integrals. 
Using such techniques, we were able to obtain expressions for the all-loop expansion of the integrated correlator in the leading and next-to-leading planar limit. We further fully resummed the perturbative expansion in terms of exact functions of the 't Hooft coupling $\lambda$, which we then expanded at strong coupling, while keeping track of their resurgent properties.

Furthermore, we explored the large-$N$ fixed-$\tau$ regime, and proposed the $SL(2,\mathbb Z)$ completion of the perturbative results. The proposal is supported by explicit instanton calculations. In particular, we computed the one-instanton contribution for fixed values of $N$ and we extrapolated the result to the large-$N$ fixed-$\tau$ regime. These contributions nicely agree with the proposed $SL(2,\mathbb Z)$ completion in terms of the non-holomorphic Eisenstein series. The $SL(2,\mathbb Z)$ completion also agrees with the results of the D3-brane/graviton scattering results in string theory in flat space.

These achievements leave a long list of open questions and possible research directions to be addressed in the future, both from the pure field theory side and from the holographic dual picture.

First of all, there are some technical points about the matrix model computation that would require further analysis to go beyond the order we consider in this paper. The recursive methods described in appendix \ref{app:recursion} are valid at finite and parametric $N$. Hence, it would be very interesting to develop a systematic methodology to treat the recursive formulas order by order in $N$, therefore allowing us to study the higher orders in the large-$N$ expansion of the integrated correlator. It would be even more interesting to obtain some recursive formula for the $\vev{\cO_2\cO_2 \cD \cD}$ integrated correlator that is the analogue of the Laplace-difference equations discovered for many different classes of integrated correlators of trace operators \cite{Dorigoni:2021bvj,Paul:2022piq,Brown:2023cpz,Brown:2023why,Paul:2023rka}.

Similarly, using the same recursive methods, it would be extremely interesting to have full analytic control over the instanton computations at large-$N$ (at least for low instanton corrections), going beyond the extrapolation performed in section \ref{sec:sl2z}. This technical advance could then be applied also to different observables and different models.

Another immediate perspective is associated with the direct application of supersymmetric localisation results to constrain the un-integrated four-point correlator $\vev{\cO_2\cO_2 \cD \cD}$ both at weak and strong coupling and the dual scattering amplitude, in analogy with the previous results for integrated correlators of trace operators. As mentioned in section \ref{sec:5} and further discussed in appendix \ref{app:loops}, our perturbative expansion coming from localisation can be used as constraints for Feynman diagram computations at high-loop orders, even beyond the proposed three-loop result from \cite{Jiang:2023uut}; this becomes particularly fruitful by combining with other methods like the analytic bootstrap. 

As discussed in section \ref{sec:sl2z}, at strong coupling the four-point correlator $\vev{\cO_2\cO_2 \cD \cD}$ can be interpreted as a  D3-brane/graviton scattering process in $AdS_5 \times S^5$. In the case of $\langle \cO_2 \cO_2 \cO_2 \cO_2 \rangle$, by understanding the analytic structures of corresponding Mellin amplitudes, it was possible to determine the Mellin amplitudes for the first few orders in large-$N$ fixed-$\tau$ expansion \cite{Chester:2019jas,Chester:2020vyz}, that reproduce exactly the known results in the flat-space limit \cite{Green:1997tv,Green:1998by, Green:2005ba}.  It would be very interesting to reproduce the flat space D3-brane/graviton amplitudes given in \cite{Bachas:1999um, Green:2000ke, Basu:2008gt, Lin:2015ixa}  in a similar manner.

In the context of more general four-point functions, one could consider the $\vev{\cO_p \cO_p \cD \cD}$ correlator, where $\cO_p$ are higher dimension trace operators dual to KK modes, and study whether they are related to $\vev{\cO_2\cO_2 \cD \cD}$ by some hidden ten-dimensional symmetry both for the un-integrated correlators \cite{Caron-Huot:2018kta, Caron-Huot:2021usw} and for the integrated correlators following ideas of \cite{Brown:2023zbr}. 

Furthermore, it is worth discussing the possible extension to $\cN=2$ superconformal field theories. Thanks to the connection with localisation in $\cN=2^*$ theory, several tools from the integrated correlator literature derive from and can be applied to $\cN=2$ SCFTs. Some results on integrated correlators in $\cN=2$ theories have been recently been obtained in \cite{Chester:2022sqb,Behan:2023fqq,Billo:2023kak}. Besides, there are no results for determinant operators in $\cN=2$. These would be extremely useful to shed some light on holographic properties of $\cN=2$ SCFTs, especially for $\cN=2$ superconformal QCD theory (with $SU(N)$ gauge group and $2N$ fundamental hypermultiplets). A first attempt would be the computation of the extremal correlator $\vev{\cD \bar\cD}_{\cN=2}$.

Let us also illustrate some intriguing connections with other classes of integrated correlators, in the presence of a conformal defect. After the planar interpretation of the $\cD\cD$ pair as the giant graviton D3-brane, the integrated correlator $\cC_\cD$ is reminiscent of the $\langle \mathbb{L}\, \cO_2 \cO_2 \rangle$ integrated correlator, recently studied in \cite{Pufu:2023vwo, Billo:2023ncz} (see also \cite{Gimenez-Grau:2023fcy}), where $\mathbb{L}$ is in general a line defect, dual to a $(p,q)$-string extended in AdS. The two examples for $\mathbb L$ explicitly computed from localisation in \cite{Pufu:2023vwo, Billo:2023ncz} are the Wilson and 't Hooft loop. In the holographic picture, the $\langle \mathbb{L}\, \cO_2 \cO_2 \rangle$ correlator describes the scattering of two gravitons off an extended $(p,q)$-string. A $(p,q)$-string transforms under $SL(2,\mathbb{Z})$ in a non-trivial way (in field theory, for example, a Wilson line is mapped to a 't Hooft line under the inversion of $SL(2,\mathbb{Z})$ transformation), therefore unlike the integrated correlator $\cC_\cD$ considered in this paper, the observable $\langle \mathbb{L}\, \cO_2 \cO_2 \rangle$ cannot be expressed in terms of modular invariant functions\,\footnote{See \cite{Pufu:2023vwo} for further discussion on modular properties and explicit non-perturbative instanton contributions to $\langle \mathbb{L}\, \cO_2 \cO_2 \rangle$ after integrating out the spacetime dependence.}. It will be interesting to further study the similarities and differences between these two observables in $\mathcal{N}=4$ SYM, and to investigate possible extensions to other conformal defect set ups.

Finally, we would like to mention the worldsheet interpretation of \cite{Jiang:2019xdz,Jiang:2019zig}, where the planar insertion of the $\cD\cD$ pair is associated with a boundary state on the worldsheet, and the three-point functions with non-BPS operators $\vev{\cD\cD \cO_{{\rm non-BPS}}}$ are written as an overlap of a boundary state with closed string states. In this picture, the non-BPS three-point function has been computed at any coupling in the large-$N$ limit by using a worldsheet $g$-function method coming from integrability. The non-BPS three-point functions enter in the un-integrated $\vev{\cO_2\cO_2 \cD \cD}$ four-point function as its OPE limit (and this method was used to compute the un-integrated four-point function at one- and two-loops). Similarly the integrated correlator $\cC_{\cD}$ can be used as an input for the un-integrated correlator, giving access to infinite sums over the BPS sector. It would be interesting to test this interplay between integrability and localisation even for other sets of observables.

\vskip 1cm
\noindent {\large {\bf Acknowledgments}}
\vskip 0.2cm
We would like to thank Marco Bill\'o, Frank Coronado, Robert de Mello Koch, Daniele Dorigoni, Zhihao Duan, Marialuisa Frau, Alessandro Georgoudis, Michael Green, Paul Heslop, Yunfeng Jiang, Shota Komatsu, Alberto Lerda, Sanjaye Ramgoolam, Rodolfo Russo, Oliver Schlotterer, Yu Wu, and Yang Zhang for many useful discussions and comments on the draft.   

A.B. is supported by a Royal Society funding RF$\backslash$ERE$\backslash$210067. C.W. is supported by a Royal Society University Research Fellowship,  URF$\backslash$R$\backslash$221015. F.G. and C.W. are supported by a STFC Consolidated Grant, ST$\backslash$T000686$\backslash$1 ``Amplitudes, strings \& duality". 

\vskip 1cm

\appendix

\section{Gaussian matrix models and recursive formulas}\label{app:recursion}

In this appendix, we review the recursive techniques that are one of the most efficient tools for the finite $N$ Gaussian matrix model computations. Following closely the references \cite{Billo:2017glv,Billo:2018oog,Beccaria:2020hgy,Galvagno:2020cgq}, we write the Gaussian partition function for a $N\times N$ matrix $a$ taking values in the gauge algebra $\mathfrak{su}(N)$ as:
\begin{align}\label{eq:GaussZ_b}
    \mathcal{Z}_0=\int da~{\rm e}^{-tr a^2}\,,~~~~~ a=\sum_{b=1}^{N^2-1} a^b\,T_b\,.
\end{align}
The integration measure is normalised such that
\begin{equation}
    da = \prod_{b=1}^{N^2-1} \frac{da_b}{\sqrt{2\pi}}\,,
\end{equation}
ensuring that $\mathcal{Z}_0=1$,
and the $\mathfrak{su}(N)$ generators are normalized as: 
\begin{align}
    \tr\,T_b\,T_c =\tfrac{1}{2}\,\delta_{bc}\,,~~~~~\tr T_b = 0\,.
\end{align}
Any observables can be inserted in the matrix model as
\begin{align} \label{eq:gauss2}
    \vev{f(a)}_0 = \int da~\rme^{-tr a^2} f(a)\,,
\end{align}
and in order to preserve gauge invariance, $f(a)$ is expected to be written in terms of traces of powers of the matrix $a$, for which it is convenient to introduce the following notation:
\begin{align}\label{t_def}
    t_{\vec p} = \vev{\tr a^{p_1} \tr a^{p_2} \dots \tr a^{p_m}}_0\,,
\end{align}
where $m$ is the length of $\vec p$. Expectation values of multitrace insertions \eqref{t_def} can be evaluated at finite $N$ through the basic Wick contraction $\vev{a_b\, a_c}_0 = \delta_{bc}$, as well as the following $\mathfrak{su}(N)$ matrix reduction formulas (also called fusion/fission identities) for arbitrary $N\times N$ matrices $X_1$ and $X_2$:
\begin{align}\label{eq:fus_fiss}
    \begin{aligned}
    \tr T^b X_1 T^b X_2 &= \frac{1}{2} \tr X_1 \tr X_2 -\frac{1}{2N} \tr X_1 X_2\,, \\
    \tr T^b X_1 \tr T^b X_2 &= \frac{1}{2} \tr X_1 X_2 -\frac{1}{2N} \tr X_1 \tr X_2\,.
\end{aligned}
\end{align}
Using such relations and starting from the initial conditions
\begin{align}
	\label{tnodd}
		t_{[p]} = 0~~~\text{for $p$ odd},~~~~\text{and}~~~~ t_{[0]}=N\,,
\end{align}
we can evaluate the general expression for $t_{\vec p}$ as a rational function in $N$ from the following recursion relations:
\begin{align}
\begin{split}\label{eq:recursion_relat}
t_{[p_1,p_2,\dots,p_m]} =   &\frac{1}{2} \sum_{j=0}^{p_1-2}  \Big( t_{[j,p_1-j-2,p_2,\dots,p_m]} \Big)
		-\frac{p_1-1}{2N}\,   t_{[p_1-2,p_2,\dots,p_m]} \\ 
		&+ \sum_{k=2}^{m}\frac{p_k}{2} \,\Big(  t_{[p_1+p_k-2,p_2,\dots,\slashed{p_k},\dots,p_m]} -\frac{1}{N} \,t_{[p_1-1,p_2,\dots,p_k-1,\dots,p_m]} \Big)\,,
\end{split}
\end{align}
where ${p_1, \ldots ,\slashed{p_k}, \ldots ,p_m}$ represents the sequence of $t$ indices without the $k$-th one. Some explicit examples are given by:
\begin{align}\begin{split}
&t_{[2]}=\frac{N^2-1}{2}\,, \qquad
t_{[4]}=\frac{(N^2-1)(2N^2-3)}{4N} \, , \\
&t_{[2,2]}=\frac{N^4-1}{4}\,, \qquad
t_{[6]}=\frac{5(N^2-1)(N^4-3N^2+3)}{8N^2} \, , \\
&t_{[4,2]}=\frac{(N^2-1)(N^2+3)(2N^2-3)}{8N}\,,\qquad
t_{[3,3]}=\frac{3(N^2-1)(N^2-4)}{8N}\,.
\end{split}
\end{align}
This procedure can be easily implemented in {\tt Mathematica} and greatly simplifies the computations in the matrix model.

\section{Gram-Schmidt mixing coefficients and their large-$N$ properties}
\label{app:largeN-beta}

We will now use the matrix model recursive techniques to compute the mixing coefficients of normal-ordered operators arising from the Gram-Schmidt orthogonalisation procedure, and to obtain the large-$N$ properties of these mixing coefficients. 

As discussed in the main text, to compute correlators using the matrix model, we need to perform Gram-Schmidt orthogonalisation, as given in \eqref{eq:On_NO}, which we quote below: 
\begin{equation}
    \begin{split}
        \label{eq:On_NO2}
   \widehat{\cO}_{\vec p}(a) = \frac{\cO_{\vec p}(a)}{\cN_{\vec p}} =  O_{\vec p}(a) +  \sum_{\vec q\,\vdash q< p}\alpha_{\vec p, \vec q}(N)~O_{\vec q}(a)  \, ,
    \end{split}
\end{equation}
where the coefficients $\alpha_{\vec p, \vec q}(N)$ are determined by:
\begin{align}\label{eq:orth_procedure2}
    \vev{ \widehat{\cO}_{\vec p}(a) \,  O_{\vec r}(a)}_0 =0\,,~~~ \forall r<p ~~~\text{and}~~ \forall ~\text{partitions~of~}r \,.
\end{align}
Here the Gaussian expectation value $\langle \,\, \rangle_0$ is defined in \eqref{eq:gauss2}, and therefore we can use the recursive formulas reviewed in the previous subsection to perform the Gram-Schmidt procedure.

Some examples of the normal ordering of operators with low dimensions of $p\leq 6$ are as follows: 
\begin{subequations}\label{NormalOrdExplicit}
\begin{align}
\widehat{\cO}_{[2]} &= O_{[2]}\!  - \frac{N^2-1}{2}\,,\\
\widehat{\cO}_{[4]}& =O_{[4]}
 - \frac{2N^2-3}{ N} \,O_{[2]}+\frac{(N^2-1)(2N^2-3)}{4N}
 \,, \phantom{\bigg|}\\
\widehat{\cO}_{[2,2]} &= O_{[2,2]}-(N^2+1)\,
O_{[2]}+\frac{N^4-1}{4}\,,\phantom{\bigg|}\\
\widehat{\cO}_{[6]} &= O_{[6]}\!-\!\frac{3( 2N^2-5)}{2N}\,
O_{[4]}\!-\!\frac{3}{2}\,O_{[2,2]}\!+\! \frac{ 15(N^4-3 N^2+3)}{ 4N^2}\,O_{[2]}\phantom{\bigg|}\!\!-\!\frac{5(N^2-1)(N^4-3\, N^2+3)}{ 8N^2}\,,\phantom{\bigg|}\\
 \widehat{\cO}_{[4,2]}&=O_{[4,2]}- \frac{N^2+7}{2}\,O_{[4]}
-\frac{2N^2-3}{N} O_{[2,2]}+ \frac{ 3(2N^2-3)(N^2+3)}{ 4N} \,O_{[2]}
\cr
& ~~~ -\frac{(N^2-1)(N^2+3)(2 N^2-3)}{ 8N} \,, \phantom{\bigg|} \\
\widehat{\cO}_{[3,3]}&=O_{[3,3]}-\frac{9}{2}\,O_{[4]}+\frac{9}{2N}
O_{[2,2]}+ \frac{ 9( N^2-4)}{4N} 
\,O_{[2]}-\frac{3(N^2-4)(N^2-1)}{ 8N}\,,\phantom{\bigg|}\\
\widehat{\cO}_{[2,2,2]}& = O_{[2,2,2]}
 - \frac{3(N^2+3)}{2}\,O_{[2,2]}+
 \frac{3(N^2+3)(N^2+1)}{4}\,O_{[2]}-\frac{ (N^2+3)(N^4-1)}{8} \phantom{\bigg|}\,.
\end{align}
\hspace{-0.25cm}
\end{subequations} 
From these expressions, we can read the explicit finite-$N$ values of $\alpha_{\vec p, \vec q}$ for $p\leq 6$. Proceeding to higher-dimensional operators in a similar way, we have computed  $\alpha_{\vec p, \vec q}$ for operators with dimension $p\leq 20$. Moreover, using \eqref{beta_to_alpha} we can immediately extract the coefficients $\beta_{\vec q, \vec p}$. Below we display explicitly some mixing coefficients (up to $q=8$) for arbitrary $N$:
\begin{subequations}\label{eq:Beta_Explicit}
\begin{align}
& \beta_{[4],[2]} = 2N - {3\over N}\,,\quad \beta_{[2,2],[2]} = N^2 + 1 \,,\\
&\beta_{[6],[4]} = 3N- {15 \over 2N} \,,~~~ \beta_{[6],[2,2]} =\frac{3}{2} \,,~~~\beta_{[6],[2]} = \frac{15N^2}{4}-{45\over 4} + {45\over 4N^2} \,,  \\
&\beta_{[4,2],[4]} = \frac{N^2}{2} + \frac{7}{2} \,,~~~ \beta_{[4,2],[2,2]} =2N - {3\over N}\,,~~~\beta_{[4,2],[2]} = \frac {3 N^3} {2} + \frac {9
       N} {4} - \frac {27} {4 N}\,, \\
&\beta_{[3,3],[4]} = \frac{9}{2} \,,~~~ \beta_{[3,3],[2,2]} =-\frac{9}{2N} \,,~~~\beta_{[3,3],[2]} = \frac{9N}{4}-{9\over N}~ , \\
&\beta_{[2,2,2],[4]} = 0 \,,~~~ \beta_{[2,2,2],[2,2]} =\frac{3N^2}{2} + \frac{9}{2}\,,~~~\beta_{[2,2,2],[2]} =\frac{3 N^4}{4}+3N^2+\frac{9}{4}\,, \\
&\beta_{[8],[6]} =4N-\frac{14}{N} \,,~~~ \beta_{[8],[4,2]} = 4\,,~~~\beta_{[8],[3,3]} = 2\,, ~~~\beta_{[8],[2,2,2]} =  0\,, \notag \\ &\beta_{[8],[4]} = 7N^2+\frac{105}{2N^2}-\frac{49}{2}\,,~~~\beta_{[8],[2,2]} = 7N -\frac{21}{N}\,,~~~\beta_{[8],[2]}= 7N^3-28 N +\frac{105}{2 N} -\frac{105}{2 N^3}\,, \\
&\beta_{[6,2],[6]} = \frac{N^2}{2}+\frac{11}{2} \,,~~~ \beta_{[6,2],[4,2]} = 3N-\frac{15}{2N}\,,~~~\beta_{[6,2],[3,3]} = 0 \,, ~~~\beta_{[6,2],[2,2,2]} =  \frac{3}{2}\,,\notag \\ &\beta_{[6,2],[4]} \!=\! \frac{3 N^3}{2} \!+\!\frac{39N}{4}\!-\!\frac{135}{4N} \,,~~~\beta_{[6,2],[2,2]} \!=\! \frac{9}{2} N^2\!-\!\frac{9}{2}\!+\!\frac{45}{4N^2} \,,~~~\beta_{[6,2],[2]}\!=\! \frac{5}{2} N^4\!+\!5N^2\!-\!30\!+\!\frac{75}{2N^2} \,, \\
&\beta_{[5,3],[6]} = \frac{15}{2} \,,~~~ \beta_{[5,3],[4,2]} = -\frac{15}{2N}\,,~~~\beta_{[5,3],[3,3]} =\frac{5N}{2}-\frac{5}{N} \,, ~~~\beta_{[5,3],[2,2,2]} =  0\,,\notag \\ &\beta_{[5,3],[4]} = 15N-\frac{105}{2N}\,,~~~\beta_{[5,3],[2,2]} = -\frac{15}{2}+\frac{45}{2N^2} \,,~~~\beta_{[5,3],[2]}= \frac{15N^2}{2}-45+\frac{60}{N^2} \,, \\
&\beta_{[4,4],[6]} = 8 \,,~~~ \beta_{[4,4],[4,2]} = 4N-\frac{6}{N}\,,~~~\beta_{[4,4],[3,3]} =-\frac{8}{N} \,, ~~~\beta_{[4,4],[2,2,2]} = 0 \,,\notag \\ &\beta_{[4,4],[4]} \!=\! N^3\!+\!\frac{35N}{2}\!-\!\frac{117}{2N} \,,~~~\beta_{[4,4],[2,2]} \!=\! 4N^2\!-\!6 \!+\!\frac{27}{N^2}\,,~~~\beta_{[4,4],[2]}\!=\! 2N^4\!+\!10N^2\!-\!\frac{99}{2}\!+\!\frac{135}{2N^2} \,,\\
&\beta_{[4,2,2],[6]} =0  \,,~~~ \beta_{[4,2,2],[4,2]} = N^2+9\,,~~~\beta_{[4,2,2],[3,3]} = 0\,, ~~~\beta_{[4,2,2],[2,2,2]} = 2N-\frac{3}{N} \,,\notag \\ &\beta_{[4,2,2],[4]} \!=\!\frac{N^4}{4}\!+\!4N^2\!+\!\frac{63}{4} \,,~~~\beta_{[4,2,2],[2,2]} \!=\!\frac{5N^3}{2}\!+\!\frac{39N}{4}\!-\!\frac{81}{4N} \,,~~~\beta_{[4,2,2],[2]}\!=\! N^5\!+\!\frac{13N^3}{2}\!+\!3N\!-\!\frac{45}{2N} \,, 
\\
&\beta_{[3,3,2],[6]} = 0 \,,~~~ \beta_{[3,3,2],[4,2]} = \frac{9}{2}\,,~~~\beta_{[3,3,2],[3,3]} = \frac{N^2}{2}+\frac{11}{2}\,, ~~~\beta_{[3,3,2],[2,2,2]} = \frac{9}{2N} \,,\notag \\ &\beta_{[3,3,2],[4]} = \frac{9N^2}{4}+\frac{81}{4}\,,~~~\beta_{[3,3,2],[2,2]} =-\frac{117}{4N} \,,~~~\beta_{[3,3,2],[2]}=\frac{3N^3}{2}+\frac{3N}{2}-\frac{30}{N} \,,\\
&\beta_{[2,2,2,2],[6]} = 0 \,,~~~ \beta_{[2,2,2,2],[4,2]} = 0\,,~~~\beta_{[2,2,2,2],[3,3]} = 0\,, ~~~\beta_{[2,2,2,2],[2,2,2]} = 2N^2+10 \,,\notag \\ &\beta_{[2,2,2,2],[4]} = 0\,,~~~\beta_{[2,2,2,2],[2,2]} = \frac{3N^4}{2}+12N^2+\frac{45}{2}\,,~~~\beta_{[2,2,2,2],[2]}=\frac{N^6}{2}+\frac{9N^4}{2}+\frac{23N^2}{2}+\frac{15}{2} \,.
\end{align}
\hspace{-0.25cm}
\end{subequations} 
We have omitted the coefficients of the form $\beta_{\vec q,[0]}$, because they do not contribute to the integrated correlator, as we discussed in section~\ref{sec:4_2}. Furthermore, we note that, in the large-$N$ limit, the coefficients $\beta_{\vec q,\vec p}$ are suppressed if any of the $p_i \in \vec p$ and $q_i \in \vec q$ are odd integers.

From the above results as well as higher order terms, which we do not show explicitly here, we are able to observe the large-$N$ structures of these coefficients. We find in general the $\beta$-coefficients can be written in the following form,
    \begin{align}
    \beta_{\vec{q},\vec{p}} = \beta_{\vec{q},\vec{p}}^{(0)} \, N^{\frac{q-p}{2}+ n-m} + \beta_{\vec{q},\vec{p}}^{(1)}\, N^{\frac{q-p}{2}+ n-m-2} + \ldots \, , \label{eq:expansion2_app}
\end{align}
where  $\vec p =[p_1,\ldots,p_m]$ and $\vec q =[q_1,\ldots,q_n]$, and $p=\sum_{i=1}^m p_i$ and $q=\sum_{i=1}^n q_i$.

For the first two orders in the large-$N$ expansion of the integrated correlator, only the $\beta_{\vec{p},\vec{q}}^{(0)}$ with even $p_i, q_i$ and $q_i>0$ are relevant. From all these explicit large-$N$ expressions, we find that 
for some simple cases of $\vec{p}$ and $\vec{q}$, the leading-$N$ coefficients  $\beta_{\vec{q},\vec{p}}^{(0)}$ take very simple forms. In particular, when $\vec p =[p]$ (i.e. the single-trace contribution), we have 
\begin{align}
    \beta^{(0)}_{\vec{q}, [p]} &= 2^{\frac{p-q}{2}} \left( \prod_{i=1}^{n} C_{\frac{q_i}{2}} \right) \sum_{i=1}^{n} \frac{1}{C_{\frac{q_i}{2}}} \binom{q_i}{\frac{q_i+p}{2}} \, , 
    \end{align}
where $C_n$ are the Catalan numbers. These coefficients are relevant for the leading planar limit of the integrated correlator, as in \eqref{eq:beta0}.  Other $\beta$-coefficients that are relevant for the computation of the integrated correlator beyond the planar limit is the case where $\vec q =[q_1, q_2], \vec p =[p_1, p_2]$, for which we find, 
 \begin{align} \label{eq:b00}
    \beta^{(0)}_{[q_1,q_2], [p_1,p_2]} =&\,  \frac{2^{\frac{p-q}{2}}}{\delta_{p_1 p_2}+1} \bigg[ C_{\frac{q_2}{2}} \binom{q_1}{\frac{q_1-p}{2}} \left({q_1-p \over 2}\right) + C_{\frac{q_1}{2}} \binom{q_2}{\frac{q_2-p}{2}} \left({q_2-p \over 2} \right) \cr
    &+ \binom{q_1}{\frac{q_1-p_1}{2}}\binom{q_2}{\frac{q_2-p_2}{2}}+\binom{q_1}{\frac{q_1-p_2}{2}}\binom{q_2}{\frac{q_2-p_1}{2}}\bigg] \, ,
\end{align}
where $p=p_1+p_2$ and $q=q_1+q_2$. 
These general expressions can be easily verified against explicit results for any given $\vec p,\, \vec q$ that we displayed earlier (and we have checked the formulas for the operators up to dimension $q=20$).  More generally, one may prove these expressions by imposing the orthogonal conditions \eqref{eq:orth_procedure} order by order in the large-$N$ expansion and using the Gaussian matrix-model recursive formulas.

\subsection{Two-point function of multitrace operators from matrix model}\label{app:2pt}
With the normal ordered operators at hand, one can then compute the correlators in $\mathbb{R}^4$ using the same recursive techniques. As a simple example, we can compute the two-point functions of normal ordered operators. A general multitrace operator on the sphere \eqref{eq:def_multiTrace} can be conveniently rewritten as follows:
\begin{equation}
    O_{\vec p} = (\tr a^{\ell_1})^{k_1} \dots (\tr a^{\ell_t})^{k_t}\,,
\end{equation}
where $\ell_1\neq \dots \neq\ell_t$ (and $\ell_m\neq 1$ for $SU(N)$ gauge group) count the independent traces, $k_1,\dots, k_t$ are the multiplicities for each trace. These two sets of indices define a partition of $p$:
\begin{equation}
    \sum_{m=1}^{t} \ell_m k_m = p \,,
\end{equation}
and hence a conjugacy class $\sigma_{\vec p}$ of the symmetric group $S_p$.
After the normal ordering procedure as described in the main text, we can compute the two-point functions of general multi-trace operators in the Gaussian matrix model, again using the recursive formulas \eqref{eq:recursion_relat}. The general expression at large $N$ is:
\begin{equation}\label{eq:2pt_MM_multiT}
    \vev{\cO_{\vec p}(a) \cO_{\vec p}(a)}_0 =\cN_{\vec p}^2\,\frac{p!}{|\sigma_{\vec p}|} \left(\frac{N}{2}\right)^{p} +O(N^{p-2})\,,
\end{equation}
where the size of the conjugacy class reads
\begin{equation}\label{eq:conjug_size}
   |\sigma_{\vec p}| = \frac{p!}{\prod_{m=1}^t \ell_m^{k_m} k_m!}\,.
\end{equation}
The matrix model normalization $\cN_{\vec p}$ is introduced to be consistent with the field theory convention. In particular choosing
\begin{align}\label{eq:N_vecp_MM}
     \cN_{\vec p} = 2^{p \over 2} %\left(\frac{|\sigma_{\vec p}|}{(p-1)!}\right)^{\frac{1}{2}}\,,
\end{align}
ensures that 
\begin{align}
    \vev{\cO_{\vec p}(a) \cO_{\vec p}(a)}_0 = \cR_{\vec p}(N)\,,
\end{align}
as given in \eqref{eq:2pt_normaliz}.

\section{Partial contraction of determinant operators and three-point coefficients}\label{app:partial_contract}
We use the results from \cite{Jiang:2019xdz}, after generalising to $SU(N)$ gauge group, combined with the matrix model techniques from the previous appendices to derive the field theory three-point coefficients \eqref{eq:3pt_FieldTh} and \eqref{eq:multi-trace0}.

We first quote the result from appendix B of \cite{Jiang:2019xdz} (up to slightly different conventions for the free propagator), where the key ingredient is the ``partially contracted giant graviton'' (PCGG) formula, defined as the Wick contraction of $N-\ell$ pairs of scalar fields in the determinants, while leaving $\ell$ pairs free:
\begin{align}
        \label{eq:part_contract}
    &\cD (x_1, Y_1)\cD (x_2, Y_2)\Big|^{U(N)}_{\text{partial contr}}= \sum_{\ell=0}^{N}d_{12}^{N-\ell} \,\cG^{U(N)}_{\ell}(x_1, Y_1; x_2, Y_2)    \,, \\
    & \cG^{U(N)}_{\ell}(x_1, Y_1; x_2, Y_2)  = (-1)^{\ell}(N-\ell)! \sum_{\substack{k_1,\ldots,k_{\ell}\\ \sum_{s}s k_s=\ell}}\prod_{u=1}^{\ell}\frac{\left(-\tr \left[(Y_1\cdot\Phi(x_1) \,Y_2\cdot \Phi(x_2))^u\right]\right)^{k_u}}{u^{k_u}k_u!}\, ,  \label{eq:part_contract2}
\end{align}
where $d_{12}$ is the spacetime/R-symmetry propagator as defined in \eqref{eq:Rsymm_space_prop}. 
The analysis of \cite{Jiang:2019xdz} is valid for the $U(N)$ gauge group (as we emphasised in the above formulas), while our computation is for $SU(N)$. However, as we will show below, the $SU(N)$ correction to the partial contraction formula up to NLO order at large $N$ resides simply in an overall factor, which cancels out when we compute the normalised three-point coefficients defined in \eqref{eq:3pt_DDO}.

The difference between $U(N)$ and $SU(N)$ in free theory computations lies in the free propagator of scalar fields:
\begin{align}
  \vev{Y_1\cdot\Phi^{i_1}_{j_1}(x_1) \,Y_2\cdot \Phi^{i_2}_{j_2}(x_2)}_{\textrm{Wick}}^{U(N)} &= d_{12}\,\delta^{i_1}_{j_2}\delta^{i_2}_{j_1}\,,\label{eq:propUN} \\
  \vev{Y_1\cdot\Phi^{i_1}_{j_1}(x_1) \,Y_2\cdot \Phi^{i_2}_{j_2}(x_2)}_{\textrm{Wick}}^{SU(N)} &= d_{12}\left(\delta^{i_1}_{j_2}\delta^{i_2}_{j_1}-\frac{1}{N}\delta^{i_1}_{j_1}\delta^{i_2}_{j_2}\right)\,,\label{eq:propSUN}
\end{align}
where $i_l,~j_l=1,\dots, N$ are fundamental indices of the gauge group. The difference between \eqref{eq:propUN} and \eqref{eq:propSUN} is given by the self-contraction term $\tfrac{1}{N}\delta^{i_1}_{j_1}\delta^{i_2}_{j_2}$. When deriving the partially contracted giant graviton formula by inserting at least one self-contraction term, the combinatorial factor scales like $(N-\ell)!(N-\ell-1)!$ instead of $(N-\ell)!^2$. The details of this computation can be found in appendix B of \cite{Jiang:2019xdz}. Therefore, as long as $\ell$ is finite, which will be the case for computing the three-point function \eqref{eq:DDO_Wick}, the difference between the $U(N)$ PCGG formula from \cite{Jiang:2019xdz} and its $SU(N)$ version in the large-$N$ limit can be expressed as\footnote{The minus sign comes from the minus sign in the $1/N$ correction in the $SU(N)$ propagator as in \eqref{eq:propSUN}.}
\begin{align} \label{eq:SUvsU}
\cG^{SU(N)}_{\ell}(x_1, Y_1; x_2, Y_2)  = \left(1- \frac{1}{N}  \right)\cG^{U(N)}_{\ell}(x_1, Y_1; x_2, Y_2)  + \ldots\,, 
\end{align}
where $\cG^{U(N)}_{\ell}(x_1, Y_1; x_2, Y_2)  $ is given by \eqref{eq:part_contract2} and the ellipses denote the higher-order corrections that are irrelevant for our computations at LO and NLO in the large-$N$ limit. 

Now, we can extract the normalised three-point coefficient of two determinants with a BPS multitrace operator $\cO_{\vec p}$, which can be obtained via the Wick contraction of \eqref{eq:part_contract} with $\cO_{\vec p}$,
\begin{align} \label{eq:DDO_Wick}
   { \vev{\cD (x_1, Y_1)\cD (x_2, Y_2)\cO_{\vec p}(x_3, Y_3)}  \over \vev{\cD (x_1, Y_1)\cD (x_2, Y_2) }  }= d_{12}^{-\frac{p}{2}} {\vev{\cG^{SU(N)}_{\frac{p}{2}}(x_1, Y_1; x_2, Y_2)  \cO_{\vec p}(x_3, Y_3)} \over \cG^{SU(N)}_{0} (x_1, Y_1; x_2, Y_2)  }\,.
\end{align} 
The $SU(N)$ correction given in \eqref{eq:SUvsU} indeed cancels out for the orders we are considering, and we can simply use the $U(N)$ formula as  given in \eqref{eq:part_contract2}. 

We may compute the colour-dependent part of the Wick contraction \eqref{eq:DDO_Wick} directly in the field theory, or using the Gaussian matrix model techniques from appendix \ref{app:recursion}. The matrix model computation can be done by promoting each operator in $\cG_\ell$ in \eqref{eq:part_contract} to the equivalent matrix model normal ordered operator. We show some examples for low values of $\ell$:
\begin{align}
\begin{split}
    &\cG_1 (a) = (N-1)! \cO_{[2]}(a)\,,~~~ \cG_2 (a) = \frac{(N-2)!}{2} \Big( \cO_{[2,2]}(a) -\cO_{[4]}(a)\Big)\,, \\
&\cG_3 (a) = (N-3)! \left( \frac{\cO_{[2,2,2]}(a)}{6}+\frac{\cO_{[6]}(a)}{3}-\frac{\cO_{[4,2]}(a)}{2}  \right)\,.
\end{split}
\end{align}
Then the normalised three-point coefficients \eqref{eq:C_p_def} can be computed in the Gaussian matrix model via the following two-point functions:
\begin{equation}
    \mathfrak{C}_{\vec p} (N) = \frac{1}{N!} \vev{\cG_{\frac{p}{2}} (a)\, \cO_{[\vec p]}}_0\,,
\end{equation}
where we have used $\cG_{0}=N!$, namely the determinant two-point function for the $U(N)$ theory. Using the techniques from appendix \ref{app:recursion} we can compute $ \mathfrak{C}_{\vec p} (N) $ for a general $N$, from which we then extract the LO and NLO in the large-$N$ expansion for the single-trace operators \eqref{eq:3pt_FieldTh} and verify the large-$N$ factorisation formula \eqref{eq:multi-trace0} displayed in the main text.

\section{Determinant operators and single-particle operators}\label{app:2}

One way to write down the normal-order version of the determinant operators  is to decompose the determinant operators in terms of trace operators, and then use the properties of the normal ordering of trace operators \eqref{eq:On_NO}. This process leads to the planar identification of determinant operators with the so-called single-particle operators (SPOs) introduced in \cite{Aprile:2020uxk}. \\
We start with the decomposition: 
\begin{align} \label{dettotraces}
    \mathcal{D}(a_i) = \sum_{\vec p \, \vdash N} C^\mathcal{D}_{\vec p}\, \cO_{\vec p}   =\sum_{\vec p\, \vdash p\leq N} C^{\mathcal{D}, {\rm GS}}_{\vec p}\, O_{\vec p} \, , 
\end{align}
where we have given both the expressions for $\mathcal{D}$ in terms of normal-ordered trace operators $\cO_{\vec p}$ and the trace operators on the sphere $O_{\vec p}$. The corresponding coefficients are given by

\begin{subequations}
    \begin{align}\label{eq:coeff_det_a}
    &C^\mathcal{D}_{\vec p} =(-1)^{N+ m}\, \frac{| \sigma_{\vec p}  | }{N!}  \, , \\
     &C^{\mathcal{D}, {\rm GS}}_{\vec p} = (-1)^{p+ m} \, \frac{| \sigma_{\vec p}  | }{p!} \frac{ (N+1)^\frac{N-p}{2} \, \left(\frac{1}{2} \right)_{\frac{N-p}{2}}}{N^{\frac{N-p}{2}} } \, , \label{eq:coeff_det_b}
\end{align}
\end{subequations}
where $m$ is the length of the vector $\vec p$, and the size of the conjugacy class $| \sigma_{\vec p}  |$ associated to the partition $\vec p$ is given in \eqref{eq:conjug_size}.

As shown by \cite{Aprile:2020uxk}, in the large-$N$ limit the determinant operators are in fact equivalent to the single-particle operators. We denote an SPO with dimension $p$ as $\cS_p$, which is defined as the linear combination of trace operators that are orthogonal to all the multi-trace operators\,\footnote{In the holographic dictionary, SPOs are dual to single particle states in AdS$_5 \times S^5$, belonging to the graviton multiplet when $p=2$ and higher Kaluza-Klein modes on $S^5$ when $p>2$ \cite{Aprile:2020uxk}. Hence, multi-point correlation functions of SPOs are dual to graviton/KK modes amplitudes on AdS$_5 \times S^5$.}:
\begin{align}\label{eq:SPO_multitr_a}
\langle \cS_p \,\, \cO_{[q_1, q_2, \ldots, q_n]} \rangle = 0 \, , \quad {\rm for} \quad  n\geq 2 \, . 
\end{align}
The SPO, when expressed in terms of trace operators, is given by \cite{Aprile:2020uxk}
\begin{align}\label{eq:SPO_multitr}
   \cS_p = \sum_{\vec p\, \vdash p} C^{\cS}_{\vec p} \cO_{\vec p} \, , 
\end{align}  
with coefficients
\begin{align}
    C^{\cS}_{\vec p} = \frac{| \sigma_{\vec p}  | }{(p-1)!} \sum_{ s \in \mathcal{P} (\{ p_1, ...,p_m \} )}  \frac{(-1)^{|s|+1} (N+1-p)_{p-\Sigma(s)} (N+p-\Sigma(s))_{\Sigma(s)}}{(N)_p -(N+1-p)_p} \, ,
\end{align}
where, again, $m$ is the length of $\vec p$. The sum is over all subsets $ s$ of the set $\{p_1, ... , p_m \}$, as the powerset $\mathcal{P}(S)$ of a set $S$ is the set of all subsets of $S$ \footnote{For example, $\mathcal{P}(\{ 1,2,3 \}) = \{ \{ \} , \{1\}, \{2\}, \{3\}, \{1,2\}, \{1,3\}, \{2,3\}, \{1,2,3\} \}$. }. $|s|$ is the number of elements of the subset $s$, and $\Sigma(s) = \sum_{s_i \in s} s_i$. 

Consider the SPO with dimension $p=N$, denoted as $\cS_N$. In the large-$N$ limit, the coefficients $C^{S}_{\vec p}$ are dominated by the subset $s=\{ p_1, ...,p_m \}$, \textit{i.e.} the full set. In this case, $|s|=m$ and $\Sigma(s)=N$, and so
\begin{align}\label{eq:coeff_SPO}
     C^{S}_{\vec p \, \vdash N} \Bigg|_{N \rightarrow \infty} = \frac{| \sigma_{\vec p}  | }{(N-1)!} \frac{(-1)^{m+1} (N)_N}{(N)_N - (1)_N} \sim \frac{| \sigma_{\vec p} | }{(N-1)!} (-1)^{m+1} \, .
\end{align}
By comparing the scaling of the coefficients \eqref{eq:coeff_det_a} with \eqref{eq:coeff_SPO}, we conclude that in the large-$N$ limit, the determinant operator is identical to the SPO $\cS_N$ (up to a normalisation factor).

\section{Matching with perturbation theory at one and two loops}
\label{app:loops}
In this appendix, we compare our exact result with explicit perturbative calculations of the un-integrated four-point correlator $\langle \cO_2 \cO_2 \cD \cD \rangle$ in the literature. This correlator has been computed up in the strict planar limit to two loops \cite{Jiang:2019xdz,Jiang:2019zig}, and more recently in \cite{Jiang:2023uut} the perturbative computation has been pushed to three-loop order.  Here we show that our matrix model result is perfectly consistent with the perturbative results at the one- and two-loop orders. However, we find disagreement with the recent three-loop result of \cite{Jiang:2023uut}. We believe that the disagreement may be generated by an incorrect assumption in \cite{Jiang:2023uut} about the Feynman integral basis at three loops. We will now discuss the comparison in detail. 

We begin by reviewing the one- and two-loop results, following \cite{Jiang:2023uut}. At one loop, the correlator is given by,\footnote{We have modified the overall normalisation of \cite{Jiang:2023uut} to be consistent with our convention.}
\begin{align}
\mathcal{H}_{\cD}(U, V; \tau, \bar \tau) \big \vert_{\textrm{ 1-loop}} =- N \left( {\lambda \over 4\pi^2}  \right) {U \over V} x_{13}^2 x_{24}^2  \prod_{1 \leq i<j \leq 4} x_{ij}^2  \int {d^4 x_5 \over \pi^2}  f^{(1)}(x_i) \, ,
\end{align}
where $f^{(1)}(x_i)$ is given by, 
\begin{align}
  f^{(1)}(x_i)=   {1 \over \prod_{1 \leq i<j \leq 5} x_{ij}^2} \, .
\end{align}
At two loops, the result takes the following form,  
\begin{align}
\mathcal{H}_{\cD}(U, V; \tau, \bar \tau) \big \vert_{\textrm{ 2-loop}} = {N  \over 8} \left( {\lambda \over 4\pi^2}  \right)^2 {U \over V} x_{13}^2 x_{24}^2   \prod_{1 \leq i<j \leq 4} x_{ij}^2  \int {d^4 x_5 \, d^4 x_6 \over \pi^4 } f^{(2)}(x_i) \, , 
\end{align}
where 
\begin{align} \label{eq:f2}
 f^{(2)}(x_i) =    {p_2(x_i) - p_1(x_i) \over \prod_{1 \leq i<j \leq 6} x_{ij}^2}\, ,
\end{align}
 with the numerators given by
\begin{align} \label{eq:p1p2}
    p_1(x_i) = {1\over 16} x_{12}^2 x_{34}^2 x_{56}^2  + P_{34; 1256} \, , \qquad
    p_2(x_i) = {1\over 4} x_{16}^2 x_{25}^2 x_{34}^2  + P_{34; 1256} \, , 
\end{align}
and $P_{34; 1256}$ represents summing over permutations on $\{3,4\}$ and $\{ 1,2,5,6\}$. 

As shown in \cite{Wen:2022oky} (see also \cite{Brown:2023zbr}), the integrated correlator defined in \eqref{eq:measure} is simply given by a sum of periods (with a minus sign) associated with the integrands $f^{(\ell)}(x_i)$, which are defined as
\begin{align}
\mathcal{P}_{f^{(\ell)}} = {1\over (\pi^2)^{\ell+1}}  \int \frac{d^4x_1 \dots d^4x_{4+\ell}}{\text{vol}[SO(2,4)]}	f^{(\ell)}(x_i) \, . 
\end{align}
Therefore, 
\begin{align}
    \cC^{(0)}_{\cD}(\lambda) \big{\vert}_{\textrm{ 1-loop}} =  \left( {\lambda \over 4\pi^2}  \right) \mathcal{P}_{f^{(1)}} \, , \qquad 
        \cC^{(0)}_{\cD}(\lambda)  \big{\vert}_{\textrm{ 2-loop}}= - {1  \over 8} \left( {\lambda \over 4\pi^2}  \right)^2 \mathcal{P}_{f^{(2)}} \, .
\end{align}
At one loop, it is well-known that $\mathcal{P}_{f^{(1)}}=6\zeta(3)$ \cite{Belokurov:1983km, Broadhurst:1995km}, so we have
\begin{align}
     \cC^{(0)}_{\cD}(\lambda) \big{\vert}_{\textrm{ 1-loop}} =  \frac{3  \zeta (3)}{2 \pi ^2} \lambda \, .
\end{align}
At two loops, each term in $f^{(2)}(x_i)$ gives the same period as $20\zeta(5)$ \cite{Belokurov:1983km, Broadhurst:1995km}; there are $12$ terms from $p_2(x_i)$ in \eqref{eq:p1p2} and three terms from $p_1(x_i)$ in \eqref{eq:p1p2} but with a minus sign. Therefore, $\mathcal{P}_{f^{(2)}}=(12-3)\times 20\zeta(5)$.  Putting everything together, we find, 
\begin{align}
 \cC^{(0)}_{\cD}(\lambda)  \big{\vert}_{\textrm{ 2-loop}} =- \frac{45  \zeta
   (5)}{32 \pi ^4}\lambda ^2 \, . 
\end{align}
They agree precisely with the first two orders in \eqref{eq:CD_pert}. 

The three-loop integrand for the correlator $\langle \cO_2 \cO_2 \cD \cD \rangle$ in the large-$N$ limit was recently proposed in \cite{Jiang:2023uut}. As we mentioned in the main text, the resulting period of the three-loop integrand given in \cite{Jiang:2023uut} does not match our prediction from localisation. Explicitly, using equation (4.24) of \cite{Jiang:2023uut} (again after an appropriate normalisation to be consistent with our conventions as for one- and two-loop computations), we find that the integrated result is given by
\begin{align} \label{eq:3loops}
  {1\over 32} \left( {\lambda \over 4\pi^2} \right)^3 \times 36 \times 70\zeta(7) = \frac{315 \zeta(7) }{256 \pi^6} \lambda^3\, . 
\end{align}
To arrive at the final expression, we have used the fact that each integral appearing in 
the three-loop correlator has the same period, which is $70\zeta(7)$ \cite{Wen:2022oky}. 
The result given in \eqref{eq:3loops} is indeed different from what we obtained from localisation computation, \textit{i.e.} $ \tfrac{595 \zeta(7)\lambda^3}{512 \pi^6}$, as shown in \eqref{eq:CD_pert}.\footnote{We were informed by Frank Coronado that the three-loop result given in \cite{Jiang:2023uut} also seems to be inconsistent with CFT data in the OPE limits.} 

In deriving the result, the authors of \cite{Jiang:2023uut} made a critical assumption about the Feynman integrals which can appear in the planar limit of the correlator $\langle \cO_2 \cO_2 \cD \cD \rangle$. In particular, it was assumed that the Feynman integrals appearing in $\langle \cO_2 \cO_2 \cD \cD \rangle$ are of the same type as the Feynman integrals for four-point correlators of half-BPS operators $\langle \cO_2 \cO_2 \cO_p \cO_p \rangle$, with $\cO_p$ being single-trace operators with fixed conformal dimensions $p$ \cite{Chicherin:2015edu}. We believe this may not be a correct assumption, since, unlike operators of fixed dimensions, the determinant operators $\cD$ do participate in the large-$N$ expansion, and we expect more general conformal integrals that were omitted in \cite{Jiang:2023uut} to appear in perturbative computations.\footnote{We would like to thank the authors of \cite{Jiang:2023uut}, Frank Coronado and Shota Komatsu for helpful discussions on this point.} The most general conformal integrals relevant to this correlator at three loops can be extracted from equations (4.2) and (4.3) of \cite{Eden:2012tu} by breaking the full $S_7$ permutation symmetry to $S_2 \times S_5$. It would also be very interesting to compute the three-loop contribution from standard Feynman diagram techniques, for example by using Feynman rules in twistor space as developed in \cite{Chicherin:2014uca,Caron-Huot:2023wdh}. 

%\section{Conventions}
%\label{appendix:conventions}

%\printbibliography
\bibliography{biblio}

\end{document}